\documentclass[12pt,preprint]{aastex}
\usepackage{amsmath}

\def\sun{\hbox{$\odot$}}
\def\R23{\mbox{$\rm R_{23}$}}

\def\kmsmpc{km s$^{-1}$ Mpc$^{-1}$}

\def\msun{M$_{\odot}$}

\def\Hb{\mbox{${\rm H}{\beta}$}}
\def\Ha{\mbox{${\rm H}{\alpha}$}}
\def\OIIIa{\mbox{${\rm [O\,III]\,}{\lambda\,5007}$}}

\def\OII{\mbox{${\rm [O\,II]\,}{\lambda\,3727}$}}

\def\NII{\mbox{${\rm [N\,II]\,}{\lambda\,6584}$}}

\shorttitle{zCOSMOS and SDSS SSFR - $\Sigma_{M}$ relations}
\shortauthors{Maier, C. et al.}

\begin{document}

\clearpage
\title{The  dependence of  star formation activity on stellar mass
  surface density and Sersic index in zCOSMOS galaxies at $0.5<z<0.9$ compared with SDSS
  galaxies at $0.04<z<0.08$\footnotemark[1]}

\author{C. Maier\altaffilmark{2}}

\email{chmaier@phys.ethz.ch}


\author{
S. J.~Lilly\altaffilmark{2},
G.~ Zamorani\altaffilmark{3},
M.~ Scodeggio\altaffilmark{4},
F.~Lamareille\altaffilmark{5},
T.~Contini\altaffilmark{5},
M. T.~Sargent\altaffilmark{2,16},
C.~Scarlata\altaffilmark{2,17},
P.~ Oesch\altaffilmark{2},
C. M.~Carollo\altaffilmark{2},
O.~ Le F\`{e}vre\altaffilmark{6},
A.~ Renzini\altaffilmark{7},
J.-P.~ Kneib\altaffilmark{6},
V.~Mainieri\altaffilmark{8},
%
%
S.~Bardelli\altaffilmark{3},
M.~Bolzonella\altaffilmark{3},
A.~Bongiorno\altaffilmark{9},
K.~Caputi\altaffilmark{2},
G.~ Coppa\altaffilmark{3},
O.~ Cucciati\altaffilmark{10},
S.~ de la Torre\altaffilmark{6},
L.~ de Ravel\altaffilmark{6},
P.~ Franzetti\altaffilmark{4},
B.~ Garilli\altaffilmark{4},
A.~ Iovino\altaffilmark{10},
P.~ Kampczyk\altaffilmark{2},
C. ~Knobel\altaffilmark{2},
K.~ Kova\v{c}\altaffilmark{2},
J.-F.~ Le Borgne\altaffilmark{5},
V.~ Le Brun\altaffilmark{6},
M.~ Mignoli\altaffilmark{3},
R.~ Pello\altaffilmark{5},
Y.~ Peng\altaffilmark{2},
E.~ Perez Montero\altaffilmark{5},
E.~ Ricciardelli\altaffilmark{7},
J.~D.~Silverman\altaffilmark{2},
M.~ Tanaka\altaffilmark{3},
L.~ Tasca\altaffilmark{6},
L.~ Tresse\altaffilmark{6},
D.~ Vergani\altaffilmark{3},
E.~ Zucca\altaffilmark{3},
%
%
U.~ Abbas\altaffilmark{6},
D.~ Bottini\altaffilmark{4},
A.~ Cappi\altaffilmark{3},
P.~ Cassata\altaffilmark{6},
A.~ Cimatti\altaffilmark{11},
M.~ Fumana\altaffilmark{4},
L.~ Guzzo\altaffilmark{10},
C.~ Halliday\altaffilmark{18,19},
A.M.~Koekemoer\altaffilmark{20},
A.~ Leauthaud\altaffilmark{5},
D.~ Maccagni\altaffilmark{4},
C.~ Marinoni\altaffilmark{12},
H.~ J. McCracken\altaffilmark{13},
P.~ Memeo\altaffilmark{4},
B.~ Meneux\altaffilmark{9,14},
C.~ Porciani\altaffilmark{2},
L.~Pozzetti\altaffilmark{3},
R.~ Scaramella\altaffilmark{15}}

\altaffiltext{2}{Institute of Astronomy,  ETH Zurich, CH-8093, Zurich, Switzerland}
\altaffiltext{3}{INAF Osservatorio Astronomico di Bologna, via Ranzani 1, I-40127, Bologna, Italy}
\altaffiltext{4}{INAF - IASF Milano, Milan, Italy}
\altaffiltext{5}{Laboratoire d'Astrophysique de Toulouse-Tarbes, Universite de Toulouse,
CNRS, 14 avenue Edouard Belin, F-31400 Toulouse, France}
\altaffiltext{6}{Laboratoire d'Astrophysique de Marseille, Marseille, France}
\altaffiltext{7}{Dipartimento di Astronomia, Universita di Padova, Padova, Italy}
\altaffiltext{8}{European Southern Observatory, Karl-Schwarzschild-Strasse 2, Garching, D-85748, Germany}
\altaffiltext{9}{Max-Planck-Institut f\"ur Extraterrestrische Physik,
  D-84571 Garching b. Muenchen, Germany}
\altaffiltext{10}{INAF Osservatorio Astronomico di Brera, Milan, Italy}
\altaffiltext{11}{Dipartimento di Astronomia, Universit\'a di Bologna,  via Ranzani 1, I-40127, Bologna, Italy}
\altaffiltext{12}{Centre de Physique Theorique, Marseille, Marseille,  France}
\altaffiltext{13}{Institut d'Astrophysique de Paris, UMR 7095 CNRS, Universit\'e Pierre et Marie Curie, 98 bis Boulevard Arago, F-75014 Paris, France}
\altaffiltext{14}{Universit\"ats-Sternwarte, Scheinerstrasse 1, D-81679 Muenchen, Germany} 
\altaffiltext{15}{INAF, Osservatorio di Roma, Monteporzio Catone (RM), Italy}
\altaffiltext{16}{Max-Planck-Institut f\"ur Astronomie, K\"onigstuhl
  17, D-69117 Heidelberg, Germany}
\altaffiltext{17}{California Institute of Technology, MS 105-24,
  Pasadena, CA 91125}
\altaffiltext{18}{Osservatorio Astrofisico di Arcetri, Largo Enrico Fermi 5, 50125 Firenze,
Italy}
\altaffiltext{19}{Department of Physics and Astronomy, University of Glasgow, Glasgow G12
8QQ, United Kingdom}
\altaffiltext{20}{Space Telescope Science Institute, 3700 San Martin Drive, Baltimore, MD 21218, USA}


\footnotetext[1]{Based on observations
  obtained at the European Southern Observatory (ESO) Very Large
  Telescope (VLT), Paranal, Chile, as part of the Large Program
  175.A-0839 (the zCOSMOS Spectroscopic Redshift Survey)}


\begin{abstract} 
In order to try to understand the internal evolution of galaxies and relate 
this to the global evolution of the galaxy population, we present a 
comparative study of the dependence of star formation rates on the average
surface mass densities ($\Sigma_{M}$) of galaxies at $0.5 < z < 0.9$ and
$0.04<z<0.08$, using the zCOSMOS and SDSS surveys respectively.  We 
derive star
formation rates, stellar masses, and structural parameters in a
consistent way for both samples, and apply them to samples that are complete down to the
same stellar mass at both redshifts.
We first show that the characteristic step-function dependence of median 
specific star formation rate (SSFR) on $\Sigma_{M}$ in SDSS, seen by
\citet{brinchmann04},
is due to the changeover from predominantly disk galaxies
to predominantly spheroidal galaxies at the surface mass density 
$log\Sigma_{Mchar} \sim 8.5$ at which the SSFR is seen to drop.
Turning to zCOSMOS, we find a similar shape for the median SSFR - $\Sigma_{M}$ 
relation, but with median SSFR values that are about 
$5-6$ times higher than for SDSS, across the whole range 
of $\Sigma_{M}$, and in galaxies with both high and low Sersic indices.  This
emphasizes that galaxies of all types are contributing, proportionally, 
to the global increase in star formation rate density in the Universe back to these 
redshifts.
The $\Sigma_{Mchar}$ "step" shifts to slightly higher values of $\Sigma_{M}$
in zCOSMOS relative to SDSS, but this can be explained by a modest
differential evolution in the size-mass
relations of disk and spheroid galaxies. 
For low Sersic index galaxies, there is little change in
the size-mass relation, as seen
by \citet{barden05}, although we suggest that this does not 
necessarily imply inside-out growth of disks, at least not in 
this redshift range.
On the other hand, there is a modest evolution in
the stellar mass - size relation for high Sersic index galaxies, 
with galaxies smaller by $\sim 25$\% at
$z \sim 0.7$.  Taken
together these produce a modest increase in $\Sigma_{Mchar}$.
Low Sersic index galaxies have a SSFR that is almost independent of
$\Sigma_{M}$, and the same is probably also true of high Sersic index
galaxies once obvious disk systems are excluded.

\end{abstract}


\section{Introduction}

One of the key unanswered questions in the study of galaxy evolution 
is what physical processes inside galaxies drive the changes in the 
star formation rates in individual galaxies that, taken together, 
produce the large
decline in the global star-formation rate density to redshifts since 
$z \sim 2$ \citep[e.g.][and references therein]{lilly96,hippel03,hopkbeac06}.
It is known that there is a strong correlation between star formation 
rate (SFR) and stellar mass in the local Universe as shown by 
numerous SDSS 
studies \citep[e.g.][]{brinchmann04,salim07,schimi07}.
This is demonstrated, for example, in
Fig.17 of \citet{brinchmann04}, although this figure also clearly
shows that at
high stellar masses the distribution of SFRs broadens significantly
and the correlation between stellar mass and SFR breaks down.  
Many studies at intermediate redshifts of the
SFR or specific SFR (SSFR) have been made as a function of the integrated
stellar mass of galaxies 
\citep[e.g.][]{bauer05,bundy06,noeske07a,noeske07b}
but these did not use information on the 
internal structural properties of the galaxies.

In contrast, studies using the local SDSS sample \citep[e.g.][]{brinchmann04}
have argued that the surface mass density
may be more important than stellar mass in regulating star formation.
Using the SDSS sample \citet{brinchmann04} found that the low specific
star formation rate (SSFR) peak is more
prominent at high $\Sigma_{M}$ than at high $M_{*}$, and therefore concluded
that the surface density of stars
is more important than  stellar mass in regulating star formation.
In a follow-up study, \citet{kaufm06} found that the total spread 
in SSFR reaches a maximum at a
characteristic stellar surface mass density  $\Sigma_{Mchar}$, and
interpreted this as a qualitative change in the
distribution of star formation histories above and below $\Sigma_{Mchar}$.
The behaviour of the SSFR with $\Sigma_{M}$ follows a smoothed ``step-function''
dropping substantially at a characteristic $\Sigma_{Mchar}$.

This observation and the resulting discussion therefore 
motivates the current investigation.  Using the HST/ACS images of 
the COSMOS field, plus star-formation rate information from emission lines
measured in large numbers of zCOSMOS spectra
we can study the changes that have occured in the SSFR-$\Sigma_{M}$ relation between
redshifts approaching $z \sim 1$ and the present epoch, as sampled by the SDSS studies,
provided we can select comparable samples at the different redshifts, e.g. above 
a certain integrated stellar mass.

The goal of this paper is therefore to obtain clues about the links between the
internal evolution of galaxies, in particular
the build-up of stellar mass, 
and the global changes that are seen in the population of galaxies as a whole, by
studying this SSFR-$\Sigma_{M}$ relationship at significantly earlier
epochs, 
$0.5 < z < 0.9$.

This work represents an improvement over an earlier study of the SSFR versus stellar 
mass surface
density in
the COSMOS field by \citet{zamoj07} by using 
secure spectroscopic redshifts, by deriving SFRs from emission line 
fluxes rather than ultraviolet continuum luminosities, by carefully constructing 
compatible
mass-complete sub-samples for both zCOSMOS and SDSS, by computing half-light radii
in a consistent way, at the same rest-frame wavelength, and by exploring
the contributions of galaxies with different Sersic indices to the overall
SSFR - $\Sigma_{M}$ relation.

By focussing on internal properties of galaxies, this paper complements a number of 
other studies of the evolution of galaxies over the $0 < z < 1$ redshift range that are
being carried
out using the first 10,000 spectra from the zCOSMOS redshift survey
\citep[e.g.][]{caputi08a,caputi08b,mignoli08,silverm08}.
These other papers are 
more focussed on the integrated
properties of galaxies and on the variation of these with the external galaxian environment.

This paper is organized as follows:
Sect.\,2  describes our sample selection and the derivation of
the relevant physical quantities for both 
zCOSMOS ($0.5<z<0.9$) and  SDSS ($0.04<z<0.08$) galaxies.
In Sect.\,3, we present our main results regarding the role of stellar mass and
stellar mass surface density in regulating star formation activity for 
zCOSMOS and SDSS galaxies as a function of morphology.  These include new insights
into the form of the SSFR - $\Sigma_{M}$ relation at the present epoch, 
and a comparison of the changes seen in this relation back to $0.5 < z < 0.9$.
These results are discussed in  Sect.\,3, and conclusions are
presented in the last section.
A cosmology with $\rm{H}_{0}=70$ \kmsmpc,
$\Omega_{0}=0.25$, $\Omega_{\Lambda}=0.75$ is used throughout this
paper.

%
\section{Datasets}
%

\subsection{The zCOSMOS sample at $0.5<z<0.9$} 

The Cosmic Evolution Survey \citep[COSMOS,][]{scoville07a} 
is the largest HST survey ever undertaken, imaging an
equatorial $1.64\,\rm{deg}^{2}$ field with single-orbit I-band
exposures \citep{scoville07b}.
The zCOSMOS project \citep{lilly07} is securing spectroscopic 
redshifts for large numbers of galaxies in the
COSMOS field. We now have in hand spectra of about 10\,500 $I_{AB} \le 22.5$ I-band selected galaxies over 
1.5 deg$^{2}$ of the COSMOS field, the so-called zCOSMOS-bright \emph{10k sample} \citep{lilly08}.
The zCOSMOS-bright sample is a flux-limited sample of galaxies with $I_{AB} < 22.5$ generated from 
the COSMOS HST ACS images. The sample does not have a significant 
observational surface brightness selection, but, as with all samples selected on total magnitude, there
is a selection in surface brightness at a given size. The success rate of measuring redshifts
varies with redshift and is very high (more than 90\%) between $0.5 < z < 0.9$.   
For more details about the zCOSMOS survey, we refer the reader to \citet{lilly07,lilly08}.

Spectroscopic observations in zCOSMOS-bright were acquired using
VIMOS with the $\rm{R}\sim 600$ MR grism over a spectral range over 5550-9650\AA. This enables
us to study rest-frame spectral features around 4000\AA, such as the
\OII\, emission line, to $z \sim 1.2$.  


\subsubsection{Sample selection}
\label{selectZC}

Only 10k sample galaxies with  reliable redshifts are used for this study.  Specifically,
and with reference to the scheme decribed by \citet{lilly08}, we use redshifts of confidence classes 
4, 3, 2.5, 2.4, 9.5, 9.4,
9.3, and 1.5.   These have an overall reliability, based on repeat observations and 
on overall consistency with independent photometric redshifts, of more than 99\%.

We exclude about 500 stars, the broad line AGNs,
and the few galaxies that lie outside the ACS fields, leaving us with a sample 
of 8131 galaxies at $z<1.2$ for which structural parameters
are available from the HST/ACS images from GIM2D Sersic fits, as described in detail
by
\citet{sargent07}. Automated morphological classifications using the ZEST scheme 
\citep[ZEST;][]{scarlata07} are also available.  

To have a high degree of spectroscopic completeness \citep[see Fig.\,9 in][]{lilly07} 
and for additional reasons discussed below, we further restrict the
zCOSMOS sample for this study to galaxies lying between $0.517<z<0.900$.

To avoid night sky lines problems with our chosen tracer for star formation, the [OII]
emission line, 
we exclude all galaxies for which the center of the \OII\, line is less
than 12\AA\, away from the center of a strong night sky line.
We therefore exclude  galaxies with \OII\, lines close to the strong 
night sky lines
5893\AA,  [OI]6300\AA, 6364\AA, or 6832\AA,  i.e. at redshifts
$0.5771<z<0.5852$, $0.6871<z<0.6936$, $0.7043<z<0.7108$, and
$0.8299<z<0.8363$, respectively.
We exclude galaxies in the redshift range $z < 0.5174$
to avoid problems with the [OII] line flux measurements using the automatic
software Platefit\_VIMOS \citep{lamar08} at the blue edge of the spectrum.
This way we remain with 3232 galaxies.


\subsubsection{Line measurements and AGN rejection}
\label{AGNrej}

Emission line fluxes were measured using  the automatic routine
Platefit\_VIMOS \citep{lamar08}.
After removing a stellar-component using \citet{bruzcharl03} models,
Platefit\_VIMOS completes a simultaneous gaussian fit of
all emission lines using a gaussian profile. 
If an emission line measurement has a significance
of less
than $1.15 \sigma$ \citep[this value is derived using the distribution of fake
detections, as described in][]{lamar08},
then an upper limit is calculated for the
respective line. 

For galaxies for which Platefit\_VIMOS detects all three emission
lines \OIIIa, \Hb, and \OII,  
we use the \OIIIa/\Hb\, versus \OII/\Hb\, diagram  to distinguish star formation dominated galaxies from
objects obviously
containing an active nucleus (narrow-line AGNs) with the aid of  the
empirical threshold derived using the 2dFGRS by \citet{lamar04}.
Additionally, all the X-ray detected AGN \citep{brusa07} are excluded,
regardless of their spectral properties. 

This results in a sample of 3048
galaxies at $0.5<z<0.9$, comprising
1527 galaxies  at $0.5<z<0.7$, and 1521 galaxies at $0.7<z<0.9$.
These objects are used for the analysis below.


\subsubsection{Structural parameters}
\label{morphpar}

Physical sizes are computed from the
half-light radii, $r_{1/2}$, defined to be the semi-major axis of the ellipse containing half of the total flux,
derived from the GIM2D Sersic fits of \citet{sargent07}. These GIM2D fits also
provide Sersic indices.  It should be noted that GIM2D provides a surface brightness profile for each galaxy as it would
be in the absence of the instrumental PSF.  

As described in \citet{sargent07}, extensive tests and simulations showed that
these fits slightly
underestimate the half-light radii of galaxies. Therefore, a correction was 
applied to the half-light radii, following their Fig.\,21, which depends 
on the half-light radius size
in pixels, the I-band magnitude (i.e. the signal to noise), and the Sersic index of the galaxy in
question. This correction was $\leq 10$\% for 78\% of the 3048 zCOSMOS
galaxies at $0.5<z<0.9$, and more than 20\% for just
2\% of the sample.

Although the ZEST automated morphological classification \citep{scarlata07} is
available for the entire zCOSMOS sample, it has not been applied to the 
SDSS.  
Therefore, for consistency, we decided to simply use Sersic
indices ($n$) to compare the structures (morphologies) 
of zCOSMOS and SDSS galaxies.
As is well-known \citep[e.g.][]{blanton03a, scarlata07} 
$n>2.5$ galaxies correspond mostly
to galaxies classified as early-types spheroid-dominated galaxies, while $n<1.5$ galaxies
are mostly
objects classified as late-type disk-dominated galaxies. 

In what follows, we will split the samples according to Sersic index, and loosely 
refer to these as disk- and spheroid-dominated galaxies, as late and early type etc., 
while endeavouring to keep the Sersic criteria clear.


\subsubsection{Ellipticity}
\label{ellipticity}

In addition to the Sersic indices, we also use an ellipticity
($q=1-b/a$) criterion to separate galaxies.  There are two motivations for this:

First,
\citet{maller08} have shown that 90\% of galaxies with
an axis ratio $b/a \leq 0.55$
are disk galaxies in an SDSS sample.  Therefore, eliminating 
high Sersic galaxies
that nevertheless have a low $b/a$ ratio should produce a cleaner sample of
early-type galaxies, since these highly inclined galaxies are unlikely
to be true early types.
Secondly, the low Sersic index
disk galaxies with $b/a \leq 0.55$ are
likely to be more affected by
dust extinction \citep[see, e.g.,][]{moellenh}, affecting estimates of 
stellar masses based on the spectral energy distributions,
star formation rates derived from emission lines and possibly also
their size estimates, all of which become more difficult.  Eliminating
highly inclined galaxies with low Sersic indices also therefore has 
advantages.

For both these reasons, we will often focus on the subset of galaxies 
with $b/a>0.55$ in both the zCOSMOS and SDSS samples, and will do so regardless
of their Sersic index. By doing so, we obtain a cleaner sample of early types and
eliminate the hardest to interpret late types. Applied consistently between the
SDSS and high redshift samples, neither should introduce systematic biasses.


\subsubsection{Stellar masses and surface mass density}
\label{StelMass}

Stellar masses, and the surface mass densities, must be derived from
estimating the mass-to-light ratio of the stellar population using 
multi-band photometry.  The advantages of doing this 
using the widest possible wavelength range and
extending as far as possible into the rest-frame near-infrared
are well-known.  

Unfortunately, while such data exist in the
COSMOS field, systematic near-infrared
photometry is not yet available for SDSS. Therefore, in order to achieve the 
highest possible {\it internal}
consistency for our study, we therefore derive mass estimates from purely 
rest-frame optical colors, for both SDSS and zCOSMOS, using 
the  equation given below in Eq.\,\ref{logM}.

\begin{figure}[h]
\includegraphics[width=16cm,angle=270,clip=true]{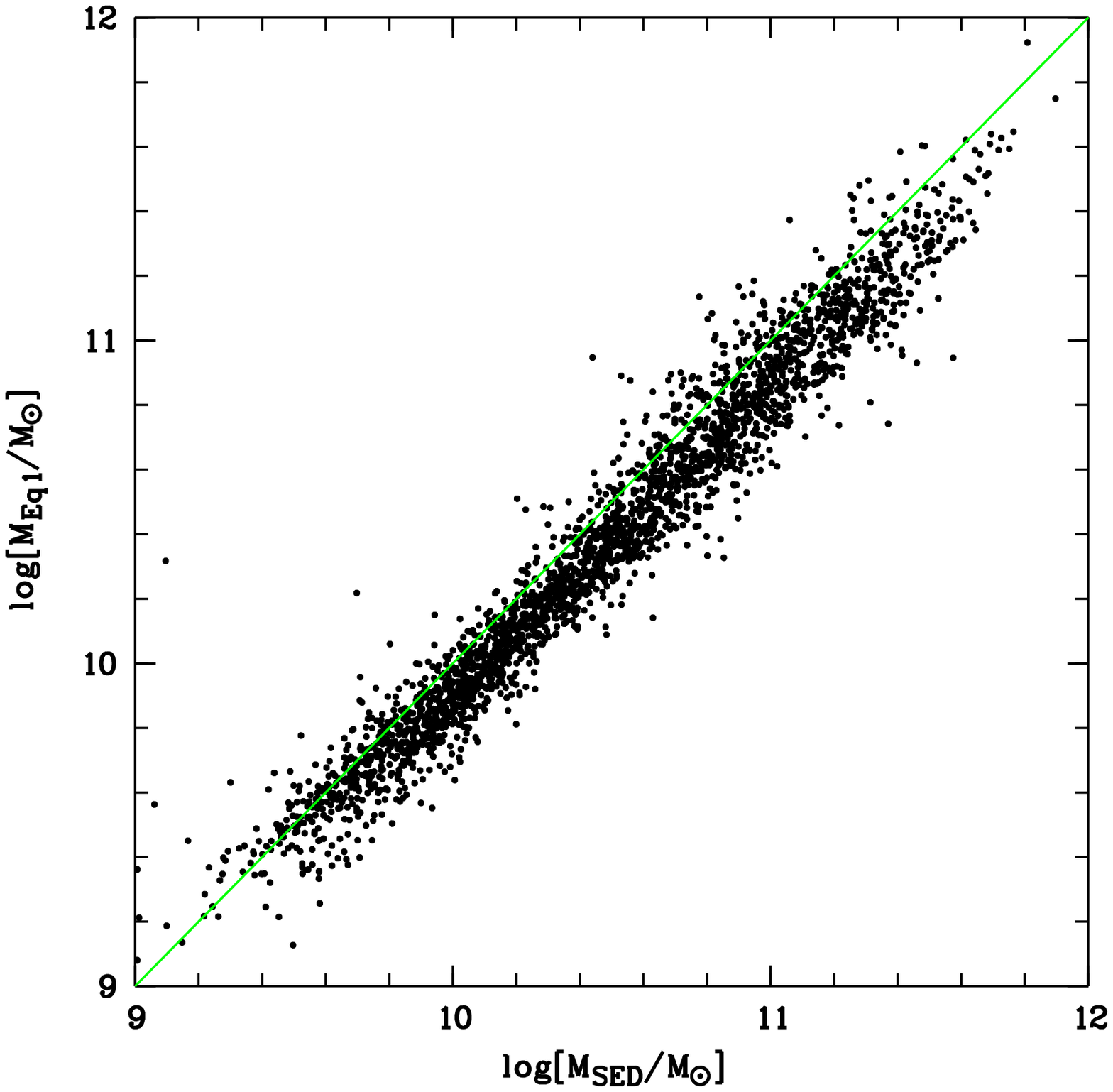}
\caption
{
\label{CompMassesSEDCol}
Comparison of  the stellar masses computed using Eq.\ref{logM} 
with masses derived using the entire COSMOS
optical to infrared SED and \citet{bruzcharl03} models by
\citet{bolzon08}. 
A good agreement with a statistical
rms of around 0.13\,dex per galaxy and an offset  (to higher SED
masses) of
0.10\,dex in the
mean is seen, reflecting typical uncertainties in derived stellar masses.
Interestingly, as described by \citet{bolzon08},  population synthesis models with TP-AGB phase
\citep{marast05}  would produce a systmatic shift of $\sim
0.1$\,dex towards lower SED masses than those shown in the figure, thus eliminating the offset
between the masses computed using Eq.\,\ref{logM} and the SED masses.
In order to achieve the 
highest possible {\it internal}
consistency for our study, we  derive mass estimates from purely 
rest-frame optical colors, for both SDSS and zCOSMOS, using 
Eq.\,\ref{logM} (c.f. discussion in Sect.\,\ref{StelMass}).
}
\end{figure}

We have however compared these stellar masses
with masses derived using the entire COSMOS
optical to infrared SED and \citet{bruzcharl03} models by \citet{bolzon08}. As shown in Fig.\,\ref{CompMassesSEDCol}
these show good agreement with a statistical
rms of around 0.13\,dex per galaxy and an offset  (to higher SED
masses) of
0.10\,dex in the
mean, reflecting typical uncertainties in derived stellar masses.
Interestingly, as described by \citet{bolzon08},  population synthesis models with TP-AGB phase
\citep{marast05}  would produce a systmatic shift of $\sim
0.1$\,dex towards lower SED masses than those shown in Fig.\,\ref{CompMassesSEDCol}, thus eliminating the offset
between the masses computed using Eq.\,\ref{logM} and the SED masses.

Stellar masses are derived  using the relation between 
rest-frame U-B and B-V colors and mass-to-light ratio (M/L), using the
Eq. (1) from \citet{lin07}, which corrects the M/L for 
evolution, and accounts for variations in the mass-to-light ratio
with color.  The relation from \citet{lin07} is modified as follows, and applied to
both SDSS and zCOSMOS data sets:

\begin{equation}
\label{logM}
\begin{split}
log(M_{*}/M_{\odot}) =  0.4 \cdot (5.48 - M_{B} - 0.10) + 1.737 \cdot
(M_{B}  - M_{V} + 0.1)\\
 + 0.098 \cdot (M_{U} - M_{B} - 0.81)  - 0.130 \cdot (M_{U} - M_{B} -
 0.81)^{2}\\
 - 0.268 \cdot z - 1.003 + \rm{log}(1.7)
\end{split}
\end{equation}

The difference between Eq. (1) in  \citet{lin07}, which uses Vega
magnitudes, and our Eq.\,\ref{logM} is due to the  transformation onto the AB system, 
using the conversions given in Table 1 in
\citet{willmer06}.
Finally, since our estimate of the SFR from
\citet{moust06}
is based on a Salpeter IMF \citep{salpeter}, and the stellar masses
computed by \citet{lin07} assume a Chabrier IMF \citep{chabrier03}, we convert the
latter masses to a Salpeter IMF
statistically, using the conversion factor of 1.7 \citep{pozzetti07}. This accounts for 
the last term in Eq.\,\ref{logM}.

For zCOSMOS galaxies, 
$M_U$,  $M_{B}$, and $M_{V}$ are Johnson rest-frame absolute magnitudes in the AB system
and were calculated from the best-fitting continuum template for each galaxy. This was obtained by
fitting the multi-wavelength COSMOS spectral energy distribution (SED)
that is available
for each object \citep{capak07, tanigu07}, using the ZEBRA photo-z code \citep{feldmann06} 
at the known, spectroscopic, redshift.   To achieve internal consistency with the 
size measurements, the SEDs were previously 
normalized to the actual (half-light) $I$-band magnitude measured from the 
GIM2D fits. 

\begin{figure}[h]
\includegraphics[width=9cm,angle=270,clip=true]{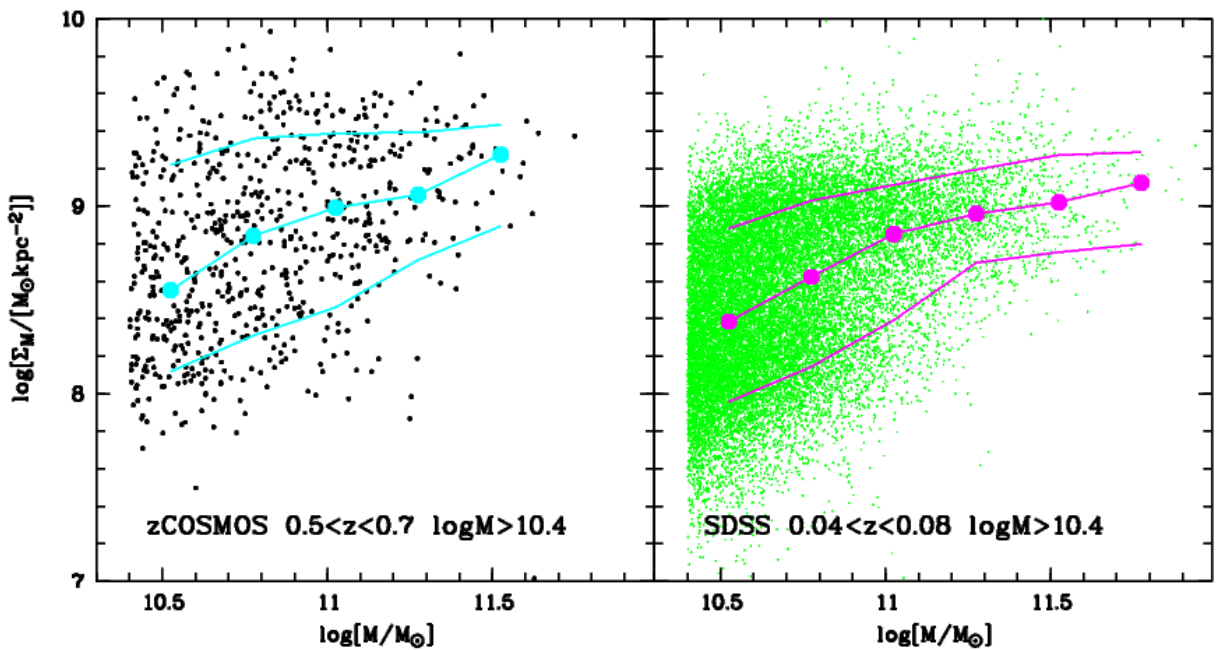}
\caption
{
\label{Mass_Marea}
Comparison between stellar mass surface density $\Sigma_{M} $ and stellar mass for  zCOSMOS
and SDSS mass-complete samples (c.f. Fig.\ref{SSFR_MassBC03}).
The median $\Sigma_{M}$ values in different stellar mass bins are shown
as filled  circles, and 16th and 84th percentiles 
as solid  lines.
A slight increase of $\Sigma_{M} $
with stellar mass, but with a large scatter, can be seen for both
zCOSMOS and SDSS samples of relatively massive galaxies (c.f.  Sect.\ref{StelMass}). 
}
\end{figure}

Finally, surface mass densities were computed using
log$\Sigma _{M}/[M_{\odot} kpc^{-2}]=log(M_{*}/M_{\odot} )- log[2 \pi (r_{1/2}/kpc)^{2}]$, 
for both the zCOSMOS and SDSS galaxies.  Further details on the
calculation of half light radii for the SDSS sample are given in Sect.\,\ref{SDSSsample}.
Since many studies of galaxy evolution at $z < 1$ are based
on stellar masses, in Fig.\,\ref{Mass_Marea} we show, for the benefit
of the reader,
 the comparison between $\Sigma_{M} $ and stellar mass for the zCOSMOS
and SDSS mass-complete samples (see below the discussion of the mass
completness).
A slight increase of $\Sigma_{M} $
with stellar mass, but with a large scatter, can be seen for both
zCOSMOS and SDSS samples of relatively massive galaxies.


\subsubsection{Star formation rates}

One of the most reliable and well-calibrated star formation rate (SFR) indicators is the
\Ha\, emission line, 
but the drawback for many high redshift studies is that \Ha\, is redshifted out of the optical window beyond
$z\sim0.5$, and its measurement requires very time-consuming  near-infrared
spectroscopy for small samples of galaxies \citep[see e.g.][]{maier05, maier06}.
Therefore, the strongest emission feature in the blue, the
\OII\, forbidden-line doublet, is usually used at $z>0.5$ as a tracer of
the SFR.
We use the \OII\, to SFR calibration of \citet{moust06} for this
study, for both the zCOSMOS objects at $0.5<z<0.9$ and the SDSS galaxies
at $0.04<z<0.08$.
%
The two dominant sources of scatter in optical SFR diagnostics, dust
extinction and metallicity, both correlate strongly with L(B) \citep[see e.g. 
Fig.\,16 in][]{moust06}, and this motivated their 
empirical calibration of SFR in terms of the  \OII\, luminosity.
This calibration was developed by tuning [OII]-derived star formation
rates to match those based on extinction-corrected \Ha\, using
observations of nearby galaxies. 
It might be a concern that the extinction changes at intermediate
redshift, but
 our own observations of \Ha\, in a limited sample
of CFRS galaxies at $0.5<z<0.9$ \citep{maier05} indicate that neither average
extinction nor average metallicity have greatly changed for most galaxies at this
redshift.
Moreover, \citet{moust06} showed their calibration 
to be effective also at intermediate redshifts.

In detail, the conversion of \OII\, luminosity to SFR is undertaken as follows.
First, because the slits in the VIMOS masks have a width of 1\,arcsec, an 
aperture correction to the \OII\, line flux is applied before computing the
SFRs.
Each zCOSMOS spectrum is convolved with the ACS I(814) filter and then
this magnitude is
compared with the I-band magnitude of the GIM2D fits of \citet{sargent07}.
The difference between the two magnitudes gives the aperture
correction factor for each spectrum. For more than 85\% of the sample,
these aperture correction factors are between 1 and 3.  This correction 
assumes that the \OII\, line flux and I-band continuum suffer equal slit 
losses.

Then, the corrected [OII] line luminosities are transformed into SFRs using a
correction factor based on the galaxy's B-band absolute magnitude, as
given by a linear interpolation of the $<R>$ values in Table 2 of \citet{moust06},
and shown in their Fig.\,19.  We use the following formula to convert the [OII] luminosity into SFR assuming a
Salpeter IMF \citep{salpeter} for both
luminous SDSS and zCOSMOS galaxies:

\begin{equation}
\label{SFR}
\rm{log}[SFR/M_{\odot}\,yr^{-1}]=\rm{log}[L_{[OII]}/(ergs\,s^{-1})]- 41  - 0.195 \cdot M_{B} - 3.434
\end{equation}

We have checked that the SFRs derived using Eq.\,\ref{SFR} are in broad
agreement with those derived  by
\citet{pozzetti08} using the entire COSMOS
optical to infrared (including K, $3.6\mu m$, and $4.6\mu m$) SED.


\subsubsection{Mass completeness of the zCOSMOS sample}
\label{masscompl}

Both zCOSMOS and SDSS are flux-limited samples.  Since the mass-to-light
ratio will depend on the stellar population, which itself will produce different
SEDs and rest-frame colors, both the SDSS and zCOSMOS samples will have a 
complicated mass-selection function.   The problem is more
acute for zCOSMOS because we can choose the SDSS redshift range
so that the SDSS sample is effectively complete for the masses
of interest.  Of course, working in terms of stellar mass instead of luminosity is
useful not only because it is closer to quantities predicted by theory,
but also because it in principle removes the evolution in luminosity due to the
aging of stellar populations.
Blue late-type star-forming galaxies generally have lower mass-to-light ratios.  Therefore, in
a given flux-limited survey such as zCOSMOS, they are detected
to smaller
masses than are red early-type galaxies, simply because the latter are fainter at a given mass, i.e.
have a higher mass-to-light ratio.
This is shown in Fig.\,\ref{zMassSDSS_zCOSMOS} (stellar mass versus redshift) and Fig.\ref{SSFR_MassBC03} 
(SSFR versus stellar mass diagram).


For this reason, if we calculate the maximum mass-to-light ratio at a given
redshift, i.e. that of 
a red passively evolving galaxy, and apply this to calculate the 
minimum mass that a
galaxy must have in order to be brighter than our flux limit, then we can safely
assume that all other galaxies with masses above this mass limit will be visible within
the survey, because they will have lower mass-to-light ratios.  

We have explored this by running a 
wide range of \citet{bruzcharl03} models, with different SFR e-folding timescales
($\tau$ from
5\,Gyrs to 0.7\,Gyrs), and with different starting redshifts ($z_{f}$ from 1.3
to 5).  For each model, we scale the model so that it has an observed $I_{AB} = 22.5$
(the flux density limit of zCOSMOS) at either $z = 0.7$ or $z = 0.9$, and 
plot the mass and SFR of
this "just-detectable" model on Fig.\,\ref{SSFR_MassBC03} as large magenta dots. 
While there is scatter because of the range of models used, these
magenta dots give a good indication of the mass completeness limit of the
sample: galaxies that occupy the diagram to the left of  the
filled circles will not be seen while those to the right will have 
been detected. 

As the mass-to-light ratio asymptotes to a limiting value the locus of the dots becomes
vertical and this limit is adopted as the overall mass completeness limit.  
These adopted mass completeness limits are shown by the green solid
vertical thick lines in the two panels of Fig.\,\ref{SSFR_MassBC03}.

Henceforth, we only consider objects above these mass limits for the two
redshift zCOSMOS redshift ranges.  At
$\rm{logM}_{*}>10.4$, we obtain a mass-complete zCOSMOS sample of 648 galaxies
at $0.5<z<0.7$, while for $\rm{logM}_{*}>10.7$ we obtain  a mass-complete zCOSMOS sample of 520 galaxies
at $0.7<z<0.9$.  As noted above and described below, the SDSS 
redshift range is chosen so that the SDSS
objects are intrinsically faint enough that we can obtain the
equivalent mass-complete samples also in SDSS.   In what follows, we will always compare
zCOSMOS and SDSS samples selected to the same mass limits.  It should be noted that
these mass limits are determined at the epoch at which the galaxy is observed, and do 
not, therefore account for any stellar mass added in between.  As we remark below,
we would expect this added mass to be modest (0.1 dex) in most cases.

%
%
\subsection{The SDSS comparison sample}
%
%
\label{SDSSsample}

\begin{figure}[h]
\includegraphics[width=16cm,angle=270,clip=true]{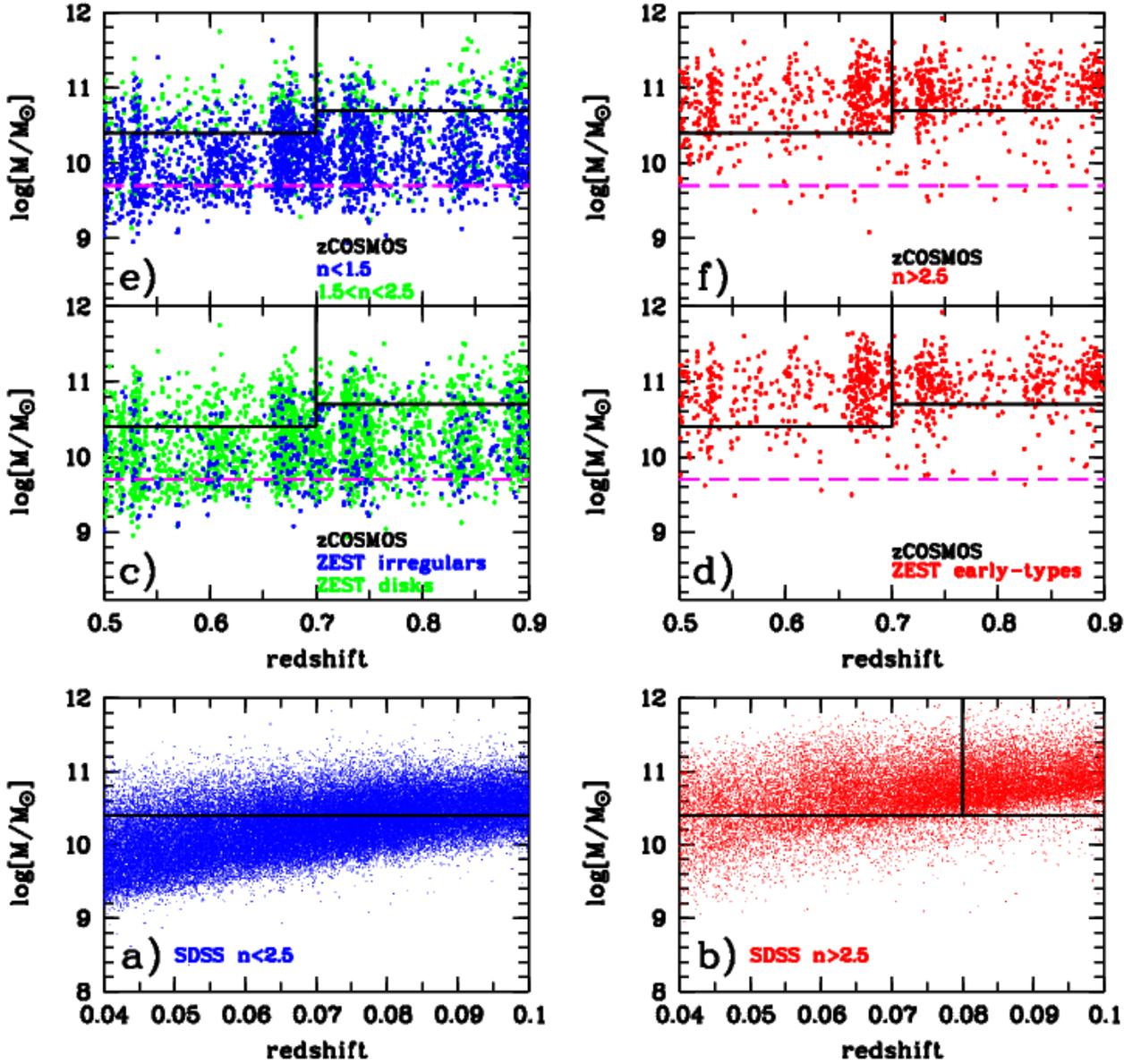}
\caption
{\label{zMassSDSS_zCOSMOS} \footnotesize 
The stellar mass as a function of redshift for  the SDSS sample (panels
a and b), and for zCOSMOS galaxies at $0.5<z<0.9$ (panels c-f) for
different Sersic indices, n. The analysis in this paper 
is based on the mass-complete samples with $\rm{log}M_{*}>10.4$ and $\rm{log}M_{*}>10.7$ 
for $0.5<z<0.7$
and $0.7<z<0.9$.
In order to be complete for SDSS galaxies (panel b) for the same mass range ($\rm{logM}_{*}>10.4$) as for
zCOSMOS galaxies, a redshift range of $0.04<z<0.08$
for the comparison SDSS sample was chosen.
}
\end{figure}

We have selected a local comparison sample of SDSS galaxies from the DR4
release, Garching repository (http://www.mpa-garching.mpg.de/SDSS/DR4/), in the redshift
range $0.04<z<0.08$. The lower redshift limit of $z=0.04$ is chosen
following the recommendation of \citet{kewley05} to reduce the systematic
and random errors in the SFRs that arise from aperture effects due to the
3\,arcsec size of the SDSS fibers. The upper limit choice of $z=0.08$
ensures mass completeness of red (high mass-to-light) SDSS galaxies 
at $\rm{logM}_{*}>10.4$,
as shown in panel b) in Fig.\,\ref{zMassSDSS_zCOSMOS}. 
We selected galaxies with Petrosian r magnitudes in the range
$14.5<r<17.77$. The bright limit is necessary because SDSS becomes
incomplete for bright galaxies of large angular size and to avoid
objects with saturated SDSS photometry, whereas the
faint limit corresponds to the nominal magnitude limit of the main galaxy
sample in SDSS.
It should be noted that duplicate objects, and SDSS galaxies on problematic  or
special plates are excluded from the SDSS sample.  This exclusion is 
based solely on position on the sky and should not 
be related to any galaxy properties.

To ensure a \emph{consistent} comparison of the physical properties of the SDSS
and zCOSMOS samples, we perform the following steps to derive stellar
masses, sizes, Sersic indices, and SFRs for SDSS galaxies:
 
a) To derive the sizes $r_{e1/2g}$ of SDSS galaxies, we start from 
the Petrosian radius containing 50\% of the Petrosian flux
in the g-band, $rpet_{50g}$. The observed g-band for SDSS galaxies at $0.04<z<0.08$
corresponds approximately to
rest-frame B-band (to be precise, effective wavelengths between 4587-4417 \AA) and 
is therefore consistent with the derivation of sizes for the zCOSMOS
sample from the observed ACS I-band images, which have effective wavelengths
of 5373{\AA} at $z=0.5$, 
4741{\AA} at $z=0.7$, and 4242{\AA} at $z = 0.9$.  Since these are rather similar, and
always above the 4000{\AA} break, any dependence on redshift (the so-called
``morphological k-correction'') should be small.
We transform $rpet_{50g}$ into half-light radii using Eq.
(6) of \citet{graham05}: $r_{1/2g} = rpet_{50g} / (1- 6 \times 10^{-6}
(rpet_{90g} / rpet_{50g})^{8.92} )$, where $rpet_{90g}$ is the
Petrosian radius containing 90\% of the Petrosian flux.  This equation
uses the concentration  $rpet_{90g}/  rpet_{50g}$ to
correct for the light outside the Petrosian aperture: this flux deficit
is 0.20\,mag in the case of de Vaucouleurs $R^{1/4}$ profile, and
0.50\,mag for an $R^{1/8}$ profile.
Then, to be consistent with the zCOSMOS $r_{1/2}$
measurements and since the SDSS Petrosian flux was measured with a circular aperture,
we transform  $r_{1/2g}$ to $r_{e1/2g}$, the semi-major axis of the
ellipse containing half of the total flux: $r_{e1/2g} = r_{1/2g} \times \sqrt{a/b}$,
where $a/b$ is the ratio between the major and minor axis of the ellipse.
We did several tests to assure the consistency of the derivation of sizes
for the zCOSMOS and SDSS samples, and the results are presented in appendix~\ref{sizetest}.
This $r_{e1/2g}$ value, the SDSS half-light radius shown in the
following in different diagrams, is then used to derive surface mass
densities as described in Sect.\,\ref{StelMass}.

b)  For the reason explained above in Sect.\,\ref{StelMass}, we use
Eq.\,\ref{logM} to derive stellar masses for the SDSS galaxies.
SDSS u, g, and r (Petrosian) magnitudes have been k-corrected
to $z=0$ according to \citet{blanton03b}, using \emph{k-correct} version 4.1.4
which is available at http://cosmo.nyu.edu/blan\-ton/kcorrect/.
U-, B- and V-band rest-frame AB magnitudes were computed from these k-corrected u, g and r magnitudes,
using the relations from Table\,2 in \citet{blantonro07}, and
stellar masses were  calculated using these values in Eq.\,\ref{logM}.
The stellar masses obtained in this way  were found to be in good
agreement with the SDSS total
stellar masses estimated from the SDSS spectra by \citet{kaufm03a}. The comparison shows
a statistical
rms of around 0.16\,dex per galaxy and an offset of 0.03\,dex in the
mean.
This is comparable to the quoted uncertainties in
the spectral measurements \citep{kaufm03a}, and with the scatter and
systematic differences between different spectral mass estimates \citep{gallazzi05}.

c) We calculate SDSS SFRs 
from the \OII\, emission line fluxes using Eq.\,\ref{SFR}, the same
equation used for zCOSMOS galaxies.
These SDSS SFRs are then  additionally corrected  using the aperture corrections given by
\citet{brinchmann04}, corrections which  should work   for $z>0.04$, as
also demonstrated by \citet{kewley05}.

d) We use the Sersic indices in the g-band from the New York University value-added
galaxy catalog (NYU-VAGC), from Sersic fits described in \citet{blanton05}.  

e) We exclude AGNs using the 
\OIIIa/ \Hb\, versus \NII/\Ha\,  diagnostic diagram, excluding objects
that satisfy  Eq. (1) of \citet{kaufm03b}.

This results in  a sample of  55230 SDSS galaxies  at $0.04<z<0.08$.
Applying the same mass cut as applied to produce the two mass-complete
zCOSMOS samples, we are left with 
21497 SDSS galaxies at $0.04<z<0.08$ with $\rm{logM}_{*}>10.4$, and 8904 SDSS galaxies at $0.04<z<0.08$ with 
$\rm{logM}_{*}>10.7$.

%
\section{Results}
%

%
\subsection{Specific star formation rates and downsizing}
%
\label{SSFR_downs}

\begin{figure}[h!]
\includegraphics[width=8cm,angle=270,clip=true]{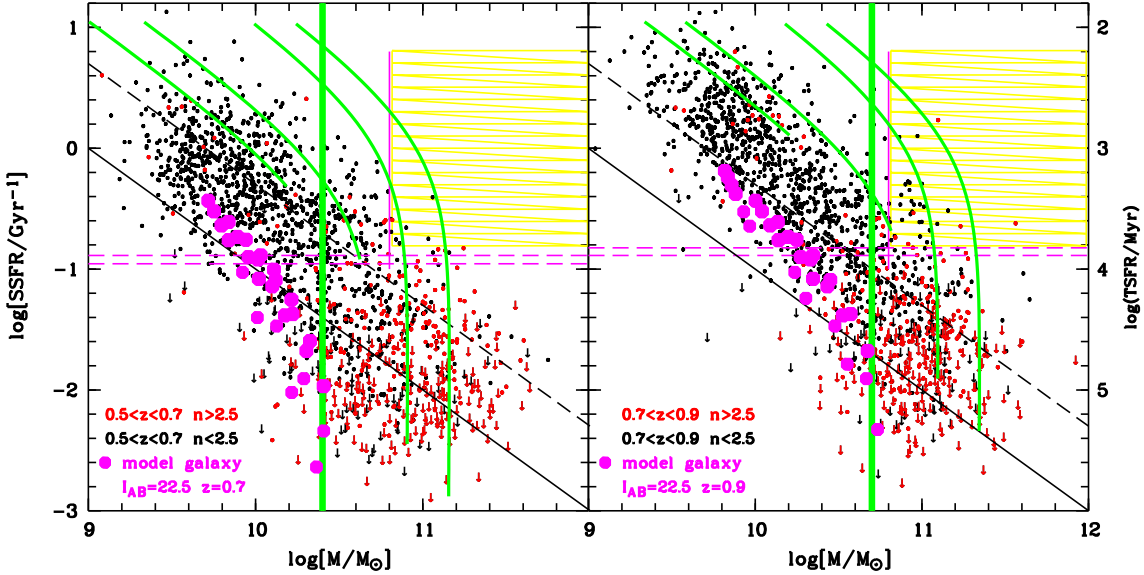}
\caption
{\label{SSFR_MassBC03} \footnotesize 
SSFR versus stellar mass as a function of Sersic
indices for the 3048
zCOSMOS galaxies  at $0.5<z<0.9$ selected as described in Sects.
\ref{selectZC} and \ref{AGNrej}.
The solid and dashed diagonal lines in each panel show star formation rates of 1 and 5
\msun/yr, respectively. 
The solid green lines show the paths of a
model galaxy to redshift $z \sim 0.6$ (left panel) and $z \sim 0.8$ (right panel) with different redshifts $z_{f}$ of the assumed onset of the star formation, and different e-folding
timescale of the star formation $\tau$: 
$z_{f}=1$ and $\tau=5$\,Gyrs, $z_{f}=2$ and $\tau=3$\,Gyrs,
$z_{f}=3$ and $\tau=1$\,Gyr, and $z_{f}=5$ and $\tau=1$\,Gyr, respectively (from left to right).
The large magenta filled circles 
show the location of a set of models
normalised to have $I_{AB} = 22.5$ (the zCOSMOS selection limit) at 
$z=0.7$ (left panel), and
$z=0.9$ (right panel), produced using
\citet{bruzcharl03} models with a wide range of different star formation histories.
These differ in having different SFR e-folding timescales
($\tau$ from
5\,Gyrs to 0.7\,Gyrs), and different starting redshifts ($z_{f}$ from 1.3
to 5).
The magenta filled circles effectively indicate the completeness limit of the
sample, in the sense that the sample 
misses galaxies that lie to the left of these. The adopted 
mass completeness limits are shown by the green vertical thick lines at
$\rm{logM}_{*}>10.4$ for $0.5<z<0.7$, and at $\rm{logM}_{*}>10.7$
for $0.7<z<0.9$.
This figure shows two
well-known phenomena:  first, the well-known downsizing effect
(illustrated by the much
higher number of galaxies in the yellow hatched region at $0.7<z<0.9$
compared to $0.5<z<0.7$),
and secondly,  the fact that galaxies with different
Sersic indices have different star formation histories, even at a 
fixed mass (c.f. discussion in Sect.\ref{SSFR_downs}).
}
\end{figure}

The star formation rate per unit stellar mass, the specific star
formation rate, is  an indicator of 
the galaxy star formation history, since 1/SSFR defines a characteristic
timescale of the stellar mass build-up.
The inverse of the specific SFR, $T_{SFR}$, is 
the time required for the galaxy to form all of its stellar mass at the
current SFR and this is shown on the right axis of Fig.\,\ref{SSFR_MassBC03}.
The ages of the universe for the two respective redshift ranges considered 
are shown as the horizontal dashed magenta lines.  Galaxies with $T_{SFR}$ higher than the 
age of the Universe at that time are relatively ``quiescent'' and must have
had a higher star formation rate in the past, while those
with $T_{SFR}$ less than the age of the Universe may be thought of as
``active'' or ``forming'', in the sense that they cannot have maintained
this high star formation rate for all of their lifetime.
If the star formation rate is constant or slowly declining, then 
``forming'' galaxies will follow a roughly diagonal track
in Fig.\ref{SSFR_MassBC03} as they age 
towards lower redshifts, whereas ``quiescent'' galaxies will 
follow a nearly vertical path downwards towards lower SSFRs
(the two models on the right in the panels of Fig.\ref{SSFR_MassBC03}).

It should be remembered that Fig.\,\ref{SSFR_MassBC03} shows 
all galaxies in the zCOSMOS samples, regardless of whether 
they belong to the mass-complete samples to the right of the 
heavy vertical lines in the two panels. This figure shows two
well-known phenomena:  first, the well-known downsizing effect is observed 
in the sense
that at $0.5<z<0.7$ very few galaxies with masses above $10^{10.8}
M_{\odot}$ (to the right of the vertical, magenta line, in the yellow,
hatched region)
have SSFRs above the magenta lines, while at $0.7<z<0.9$ there exist
several dozens of galaxies above $10^{10.8} M_{\odot}$ that are ``forming''.
Secondly, it is also obvious from the figure that galaxies with different
Sersic indices have different star formation histories, even at a 
fixed mass:
late-type galaxies with $n<2.5$ have mainly
high SSFRs, which indicate continuing star formation, and possibly a 
recent onset of star formation, whereas early-type galaxies with $n > 2.5$ have mostly 
low SSFR with $T_{SFR}$
higher than the age of the Universe,
indicating higher SFRs in the past and/or a higher redshift of
their onset of star formation. 

%
\subsection{The SDSS SSFR - $\Sigma_{M}$ relation}
\label{SDSSSFR}
%

\begin{figure}[h]
\includegraphics[width=16cm,angle=270,clip=true]{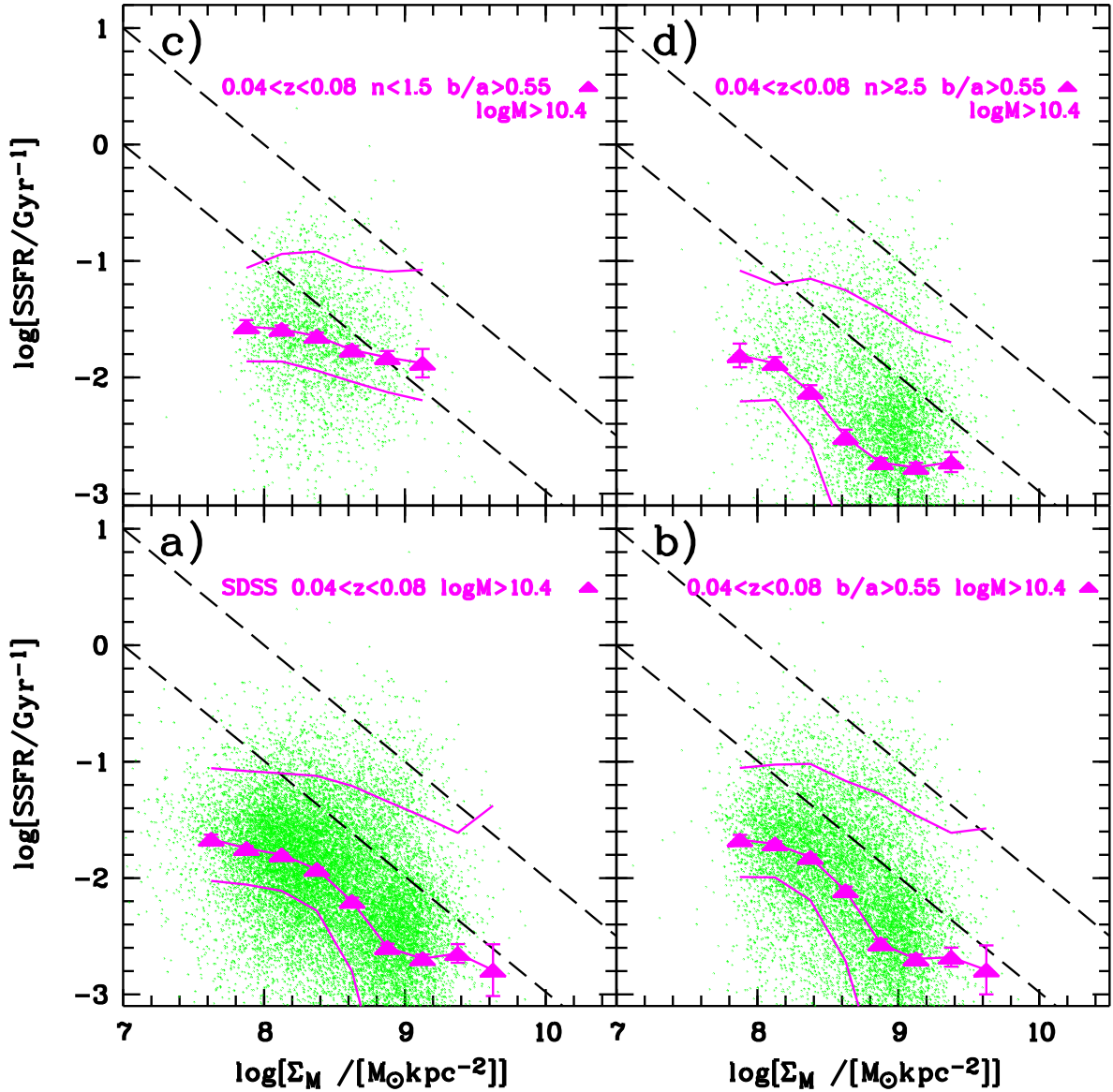}
\caption
{\label{SSFR_MareaSDSS} \footnotesize 
Specific star formation rates for 
SDSS $\rm{logM}_{*}>10.4$ galaxies at $0.04<z<0.08$  - the sample that should be directly 
comparable to the
$0.5 < z < 0.7$ zCOSMOS sample - are shown as green dots, 
the median SSFR values in different $\Sigma_{M}$ bins
as magenta triangles, and 25th and 95th percentiles \citep[as in Fig.\,9 of][]{kaufm06}
as solid magenta lines.
The diagonal dashed lines correspond to SFR surface densities
$\Sigma_{SFR} = 0.01  M_{\sun}/\rm{yr/kpc}^{2}$, and $0.1 M_{\sun}/\rm{yr/kpc}^{2}$, respectively.
Panel a)  shows all SDSS galaxies at $0.04<z<0.08$ with
$\rm{logM}_{*}>10.4$, while panel b)
shows only the SDSS galaxies with an axis ratio $b/a>0.55$ (see
\ref{ellipticity}). Panel c) shows
the low Sersic index ($n<1.5$) objects with $b/a>0.55$, and panel d)
the $n>2.5$ SDSS galaxies, again with $b/a>0.55$.
The trend seen in panel a) and b) is clearly due to the change-over of different structural types
from disk-dominated low Sersic index ($n < 1.5$) galaxies (in panel c) to bulge-dominated high Sersic ($n > 2.5$) 
index galaxies (panel d)
as the $\Sigma_{M}$ increases. 
This change-over is also clearly seen in Fig.\,\ref{fracngt25SDSS}.
}
\end{figure}

In Fig.\,\ref{SSFR_MareaSDSS}, we plot the SSFR versus $\Sigma_{M}$ diagram for the 
$0.04<z<0.08$
SDSS galaxies with $\rm{logM}_{*}>10.4$ - the sample that should be directly 
comparable to the
$0.5 < z < 0.7$ zCOSMOS sample. Individual
measurements are shown as green dots, the median SSFR values in different $\Sigma_{M}$ bins as magenta 
triangles, and, following  Fig.\,9 of \citet{kaufm06}, 25th and 95th percentiles of the
distribution of
SSFR in the respective  $\Sigma_{M}$ bin as solid magenta lines.
The error bars shown are estimates of the error in the median, and are computed
as $\Delta x = (x_{0.84} - x_{0.16}) / \sqrt{N}$, where N is the number
of galaxies in each $\Sigma_{M}$ bin, and  $x_{0.84}$ and $x_{0.16}$ denote
the 84th and 16th percentiles of the SSFR distribution.
Panel a) shows all galaxies, while panel b) shows only the 
SDSS galaxies with an axis ratio $b/a>0.55$ (see \ref{ellipticity} above).  Panel c)
shows the low Sersic index ($n<1.5$) objects with $b/a>0.55$, and panel d)
shows the $n>2.5$ SDSS galaxies, again with $b/a>0.55$.

%
In a previous study of the SSFR - $\Sigma_{M}$ relation for the SDSS sample, \citet{kaufm06} 
found evidence that the distribution of star formation histories
changes qualitatively above a characteristic surface mass density
$log\Sigma_{Mchar}\sim 8.5$.
Moreover, for higher mass SDSS galaxies, \citet{kaufm06} showed in their Fig.\,9 that the average
SSFR remains constant below  $\Sigma_{Mchar}$ and decreases at higher
surface mass densities.  This behaviour is seen in panels a) and b) of
Fig.\,\ref{SSFR_MareaSDSS}. 

\begin{figure}[h]
\includegraphics[width=16cm,angle=270,clip=true]{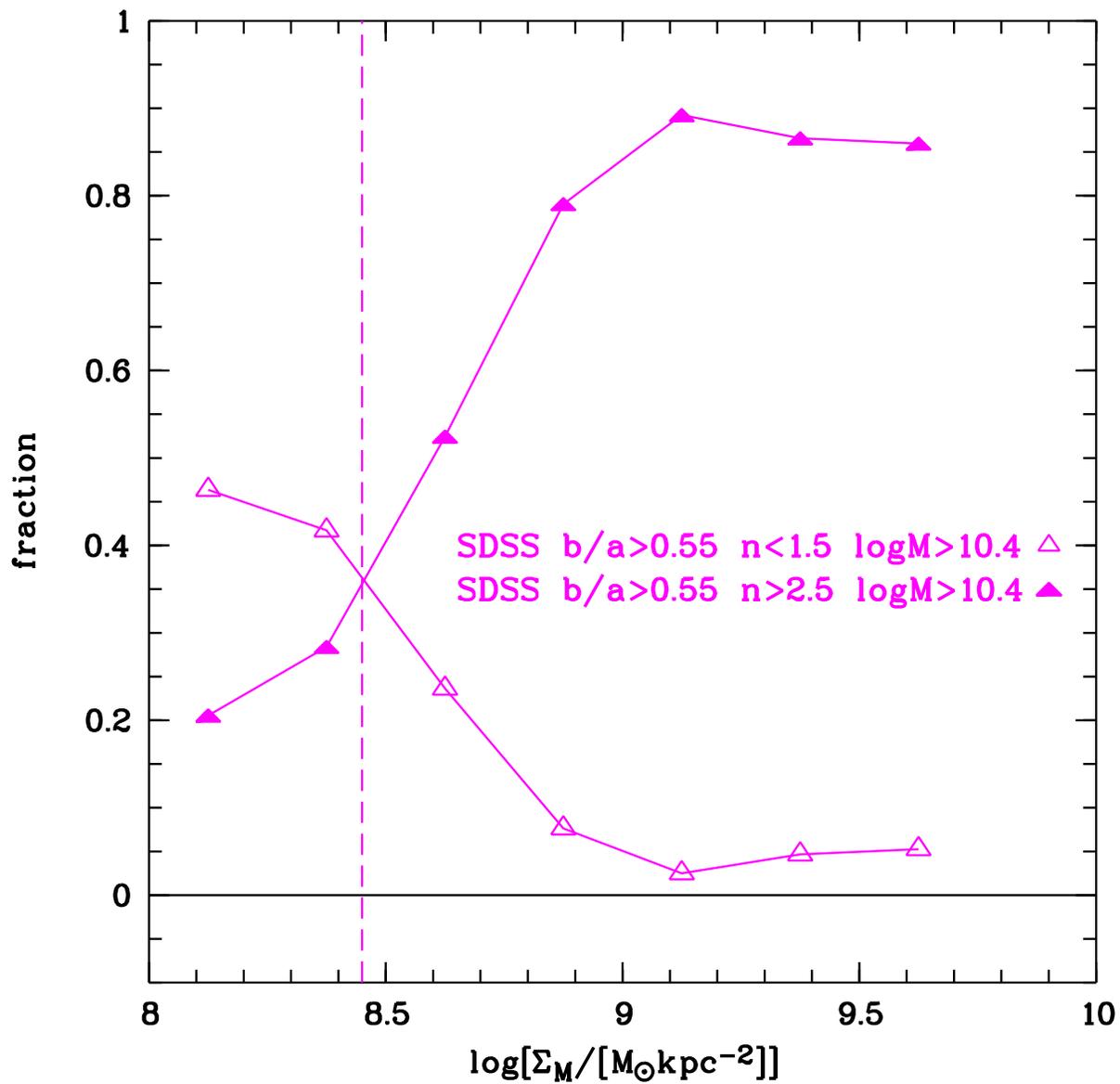}
\caption
{\label{fracngt25SDSS} \footnotesize 
The fraction of SDSS objects (with $b/a > 0.55$) with $n<1.5$ (open magenta triangles) and with
$n>2.5$ (filled magenta triangles) in the $\rm{logM}_{*}>10.4$ mass-complete sample.
The fraction of  $n>2.5$ SDSS objects shows a sharp
increase at
$log \Sigma_{Mtrans} \sim 8.45$,
very close to the characteristic surface mass density $log \Sigma_{Mchar} \sim 8.5$
found by \citet{kaufm06}.
}
\end{figure}

This trend is clearly due to the change-over of different structural types
from disk-dominated low Sersic index ($n < 1.5$) galaxies (in panel c  of
Fig.\,\ref{SSFR_MareaSDSS}) to bulge-dominated high Sersic index ($n > 2.5$) 
galaxies (panel d  of
Fig.\,\ref{SSFR_MareaSDSS})
as the $\Sigma_{M}$ increases. 
This change-over is clearly seen in Fig.\,\ref{fracngt25SDSS}, which shows
the fraction of $n<1.5$ and $n>2.5$ SDSS 
galaxies (always with
$b/a>0.55$) in
different $\Sigma_{M}$ bins (magenta open and filled triangles).
The fraction of  $n>2.5$ SDSS objects shows a sharp
increase at the point where the SSFR  abruptly changes in
Fig.\,\ref{SSFR_MareaSDSS}, i.e., at a transition surface mass density
$log \Sigma_{Mtrans} \sim 8.45$ (dashed magenta vertical line in Fig.\,\ref{fracngt25SDSS}),
close to the characteristic surface mass density $log \Sigma_{Mchar} \sim 8.5$ 
of the break in the SSFR - $\Sigma_{M}$ relation.  
%

Inspection of panel c) of Fig.\,\ref{SSFR_MareaSDSS} shows
that the median SSFR for the late-type low Sersic SDSS galaxies ($n<1.5$) 
and the plotted percentiles
are all almost
constant with $\Sigma_{M}$ - they decline very slightly with increasing $\Sigma_{M}$.  
In contrast, we see a rather 
steep drop in both the median and the 25th
percentile of SSFR for the early-type $n>2.5$ SDSS galaxies (panel d), down to a 
SSFR level 
about 5-6 times below that of the late-type galaxies, where the bulk
of the $n > 2.5$ sample resides. 
We return to the form of the SSFR$-\Sigma_{M}$ for $n>2.5$ galaxies (panel d) 
below.  In the meantime, we note that Fig.\,\ref{fracngt25SDSS} and 
panels c) and d) of Fig.\,\ref{SSFR_MareaSDSS} explain how the shape of the 
median SDSS SSFR$-\Sigma_{M}$ relation for all galaxies is the 
result of the different 
SSFR$-\Sigma_{M}$ relations for $n<1.5$ and $n>2.5$ galaxies and the 
increasing fraction of early-type $n>2.5$ SDSS galaxies
with increasing $\Sigma_{M}$ above $\Sigma_{Mtrans}$.  The large difference in the
median SSFR for early and late-type ($n > 2.5$ and $n < 1.5$) galaxies ``automatically''
makes $\Sigma_{Mchar} \sim \Sigma_{Mtrans}$.

This explanation of the overall SSFR$-\Sigma_{M}$ relation should 
not be surprising but has not, as far as we are aware, been
remarked upon previously, although \citet{schimi07} did note the
different behavior of $n < 2.5$ and $n > 2.5$ SDSS galaxies.
%

%
\subsection{The zCOSMOS SSFR - $\Sigma_{M}$ relation at $0.5 < z < 0.9$}
%
\label{zCOSMOSSSFR_SM}

\begin{figure}[h]
\includegraphics[width=14cm,angle=270,clip=true]{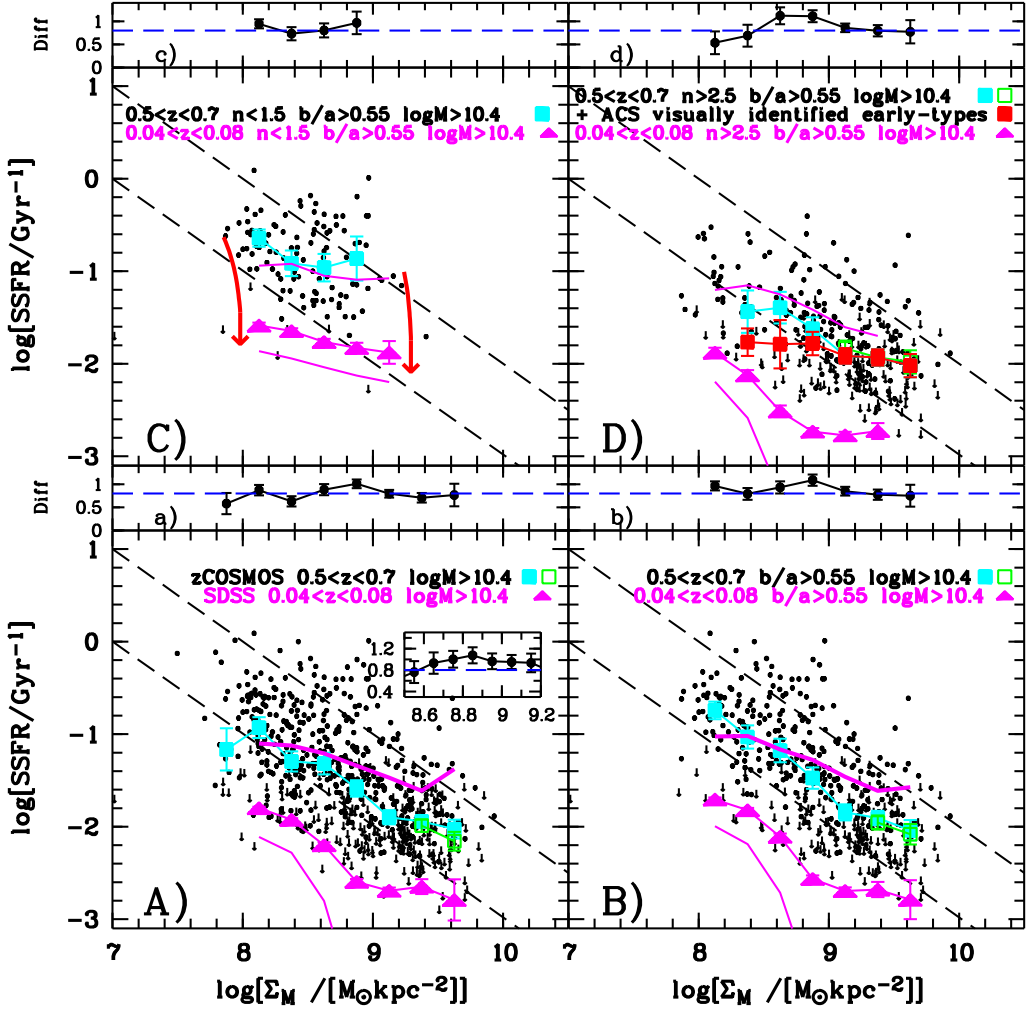}
\caption
{\label{SSFR_Mareazlt07} \footnotesize 
The specific star formation rate SSFR versus stellar surface mass density $\Sigma_{M}$
of the
zCOSMOS mass-complete sample at
$0.5<z<0.7$ (black dots), compared 
to the median and percentile curves summarizing the SDSS relations from Fig.\,\ref{SSFR_MareaSDSS}
(magenta points and lines).
The diagonal dashed lines correspond to SFR surface densities
$\Sigma_{SFR} = 0.01  M_{\sun}/\rm{yr/kpc}^{2}$, and $0.1 M_{\sun}/\rm{yr/kpc}^{2}$, 
respectively.
The SSFR median values of zCOSMOS galaxies for different $\Sigma_{M}$ bins 
are shown as cyan filled squares, while for the highest $\Sigma_{M}$ we also
show the mean SSFR values derived from co-added spectra as green open squares.
Panel A) shows all galaxies in the zCOSMOS mass-complete sample, panel
B) only galaxies with an axis ratio $b/a>0.55$, 
panel C) galaxies with $b/a>0.55$ and $n<1.5$, and panel D) galaxies
with $b/a>0.55$ and $n>2.5$.
The respective small panels above the main ones (and the insert in
panel A with finer grid in $\Sigma_{M}$) show as filled squares (circles) the difference between the median
zCOSMOS and SDSS SSFR  values in a given
$\Sigma_{M}$ bin.
For $n<1.5$ (disk) galaxies the SSFR stays roughly constant
with $\Sigma_{M}$, with the median SSFR at a given
$\Sigma_{M}$ being about six times higher at $z \sim 0.6$ than in SDSS. 
Red arrows in panel c are model galaxy tracks discussed in Sect.\,\ref{masssize}.
For $n>2.5$ (early-type) galaxies the SSFR declines with $\Sigma_{M}$ for both
zCOSMOS and SDSS galaxies, but there is again a shift to higher SSFR with
about the same factor of about five to six, and
also a modestly higher $\Sigma_{M}$ at the higher redshift.
Almost all the galaxies causing the 
upturn at low $\Sigma_{M}$ are clearly galaxies with disks, and the median
SSFR of $n>2.5$ genuine early-type
galaxies (see Sect.\,\ref{zCOSMOSSSFR_SM}) is also almost independent of $\Sigma_{M}$ (red squares in panel d). 
}
\end{figure}

\begin{figure}[h]
\includegraphics[width=16cm,angle=270,clip=true]{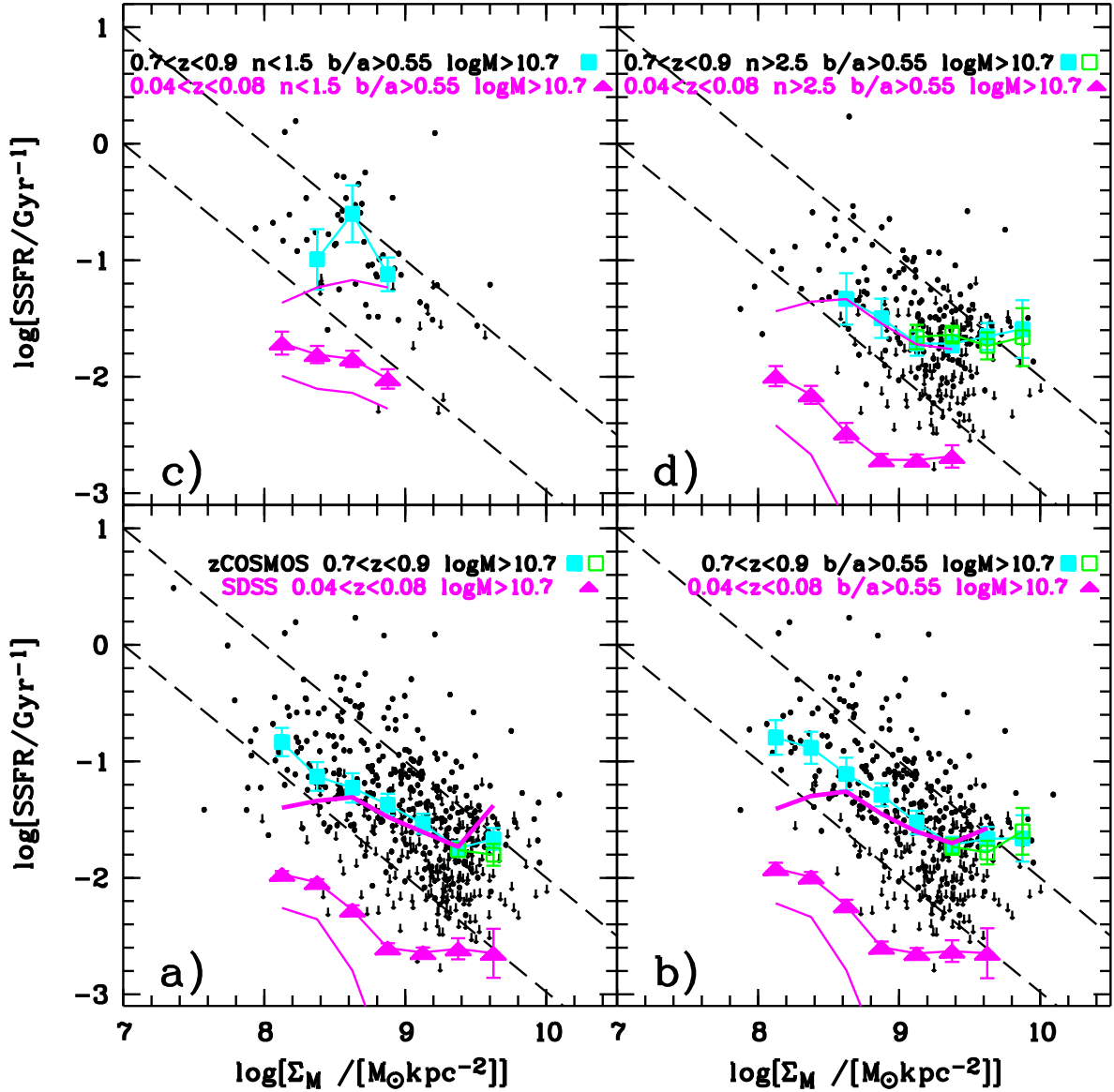}
\caption
{\label{SSFR_Mareazgt07} \footnotesize 
The specific star formation rate SSFR versus stellar surface mass density $\Sigma_{M}$
of the
zCOSMOS mass-complete sample at
$0.7<z<0.9$ (black dots), compared to SDSS (magenta symbols and lines).
Symbols are as in 
Fig.\,\ref{SSFR_Mareazlt07}, and  similar trends of the SSFR  as a
function of $\Sigma_{M}$ and
Sersic index  for $0.7<z<0.9$ zCOSMOS galaxies  are seen as for the
slightly lower redshift $0.5<z<0.7$ objects
(Fig.\,\ref{SSFR_Mareazlt07}). 
}
\end{figure}

Figures \ref{SSFR_Mareazlt07} and \ref{SSFR_Mareazgt07} show the SSFR
vs. $\Sigma_{M}$ for zCOSMOS galaxies at $0.5<z<0.9$, and the comparison
with the median and percentile curves summarizing the 
SDSS relations from  Fig.\,\ref{SSFR_MareaSDSS}.
The individual measurements for the
zCOSMOS mass-complete sample at
$0.5<z<0.7$ (Fig.\,\ref{SSFR_Mareazlt07}) and $0.7<z<0.9$ 
(Fig.\,\ref{SSFR_Mareazgt07}) are shown as black dots, and zCOSMOS median SSFR
values for different $\Sigma_{M}$ bins  as cyan, filled squares.

It should be noted that in SDSS the number of galaxies with no
measurements (upper limits) of SSFR is always lower than 50\% in all
$\Sigma_{M}$ bins we are considering, so the median SSFR in the
respective $\Sigma_{M}$ bin is always well-defined.
Unfortunately, this is not the case for the highest  $\Sigma_{M}$ bins ($log\Sigma_{M}>9$)
in the zCOSMOS
sample:
more than 50\% of the galaxies in these bins have only upper limits for 
the [OII] fluxes from Platefit\_VIMOS.
To overcome this problem, we first simply computed a median by setting
all upper limits to actual values. However, we also generated an 
"average" co-added spectrum of all the
spectra in each bin, and derived a mean SSFR from that, as follows.
%
Each individual zCOSMOS spectrum (with a spectral resolution of 2.55\AA) was
first transformed to its
rest-frame and rebinned to a common grid of 2\AA\, per pixel.
These individual rest-frame spectra were then normalized to the average
value in a featureless region of the continuum and averaged.
The [OII] flux from the co-added spectrum was transformed to a SFR using the
usual calibration, and to a SSFR using
an average mass of the galaxies in the respective
$\Sigma_{M}$ bin.
In fact, the mean SSFR value for the $log\Sigma_{M}>9$ bins
(shown as green open squares in Figs.\,\ref{SSFR_Mareazlt07} and
\ref{SSFR_Mareazgt07}) derived in this way were found to be in good
agreement
with the median values that were obtained by setting the upper limits to their maximum values
(cyan symbols).

For reasons mentioned in Sect.\,\ref{ellipticity}, we focus primarily on 
galaxies with $b/a>0.55$. Comparison of panels a and b of 
Figs.\,\ref{SSFR_Mareazlt07} and \ref{SSFR_Mareazgt07} shows that 
this does not have a big effect on the analysis.
The SSFR$-\Sigma_{M}$ relation in zCOSMOS has the same general shape as that for
the SDSS across the whole range of $\Sigma_{M}$.  However, the curve is shifted
across the full range of $\Sigma_{M}$ to
SSFR values that are about a factor of 5$-$6 higher in zCOSMOS than in SDSS.  There is 
also a small shift of about 0.1-0.2 dex 
in the characteristic $\Sigma_{Mchar}$ towards higher surface mass densities 
at higher redshifts, indicated by a localised blip in the change in SSFR
around $\Sigma_{Mchar}$.
The shift in  $\Sigma_{Mchar}$ is seen in the small panels of
Fig.\,\ref{SSFR_Mareazlt07} which show the difference between the median
zCOSMOS and SDSS SSFR values in a given $\Sigma_{M}$ bin. If the
shape of the SSFR-$\Sigma_{M}$ relation stays the same but there is a
horizontal shift in $\Sigma_{Mchar}$, then the difference between the median
zCOSMOS and SDSS SSFRs will be constant but with a localized excess in
the region over which $\Sigma_{Mchar}$ has increased. This is seen as a blip
in the small panels a, b, and d of Fig.\,\ref{SSFR_Mareazlt07}.
To illustrate more clearly the amount by which  $\Sigma_{Mchar}$ shifts
(0.1-0.2 dex) towards higher surface mass densities at higher redshifts,
we have also included in Fig.7 A an insert showing a part of panel a with a finer grid in $\Sigma_{M}$.

The rise in median SSFR is seen both in the entire sample of zCOSMOS
and SDSS galaxies (with or without the $b/a > 0.55$ cut) and in the
individual $n < 1.5$ and $n > 2.5$ subsets shown in panels c and d of the Figs.\,\ref{SSFR_Mareazlt07} and \ref{SSFR_Mareazgt07}.
This rise in SSFR is essentially the same as the rise in the global
star-formation rate density (SFRD) in the Universe as a whole to this
redshift \citep[e.g.][and references therein]{lilly96,hippel03,hopkbeac06}.
The importance of this result is that it indicates that 
galaxies of all $\Sigma_{M}$ are contributing, proportionally, to this
global evolution in the SFRD.
Although the high $\Sigma_{M}$ galaxies with 
$\Sigma_{M} > \Sigma_{Mchar}$
(mostly early type galaxies
with $n > 2.5$) 
have a SSFR that is an factor of about six lower, on average, than
the galaxies of the same mass with lower $\Sigma_{M} < \Sigma_{Mchar}$ (generally
late type galaxies with $n < 1.5$), 
the rise in SSFR by a factor of 5-6
is seen across the board.

As in SDSS, the SSFR for $n<1.5$ galaxies (panels c in Figs.\,\ref{SSFR_Mareazlt07} and \ref{SSFR_Mareazgt07}) remains roughly constant
with $\Sigma_{M}$, albeit at a level that is elevated with respect to SDSS by 
about a factor of six.  
The shape of the relation for $n>2.5$ galaxies (panel d of Figs.\,\ref{SSFR_Mareazlt07} and \ref{SSFR_Mareazgt07})
is also similar to that seen in the SDSS, with a decline in the SSFR around 
$\Sigma_{Mchar}$ and a roughly constant SSFR at higher surface mass densities.
The SSFR is again uniformly elevated across the range of $\Sigma_{M}$ 
by a factor of about five to six.

With the better resolution of the ACS COSMOS images we can now examine the galaxies
which are causing the
upturn in SSFR of $n > 2.5$ galaxies at the lower $\Sigma_{M}$ in panel d of the 
Figs.\,\ref{SSFR_Mareazlt07} and \ref{SSFR_Mareazgt07}. Most of the 
zCOSMOS galaxies with high SSFRs and lower $\Sigma_{M}$ that are responsible 
for this
upturn 
are clearly disk galaxies with a dominant
bulge. If we remove these from the sample, as indicate by the red 
points in the panel d of Fig.\,\ref{SSFR_Mareazlt07}, then we find a SSFR$-\Sigma_{M}$ relation that is
essentially flat. We suspect that the same would also hold for the $n>2.5$
SDSS sample.   Although we have had to parameterize the galaxy morphology 
in terms of a crude Sersic index, in order to achieve uniformity between SDSS
and zCOSMOS, we would predict that the form of the SSFR$-\Sigma_{M}$ would
consist of two essentially flat relations in SSFR$-\Sigma_{M}$ 
for disks and spheroids separately,
which partly overlap in $\Sigma_{M}$ and which are offset in SSFR by a
factor of five to six.  Both components appear to increase in SSFR by the same amount
to the redshifts probed by this study.

%
\subsection{The SDSS and zCOSMOS stellar mass$-$size relations}
%
\label{masssize}

\begin{figure}[h]
\includegraphics[width=16cm,angle=270,clip=true]{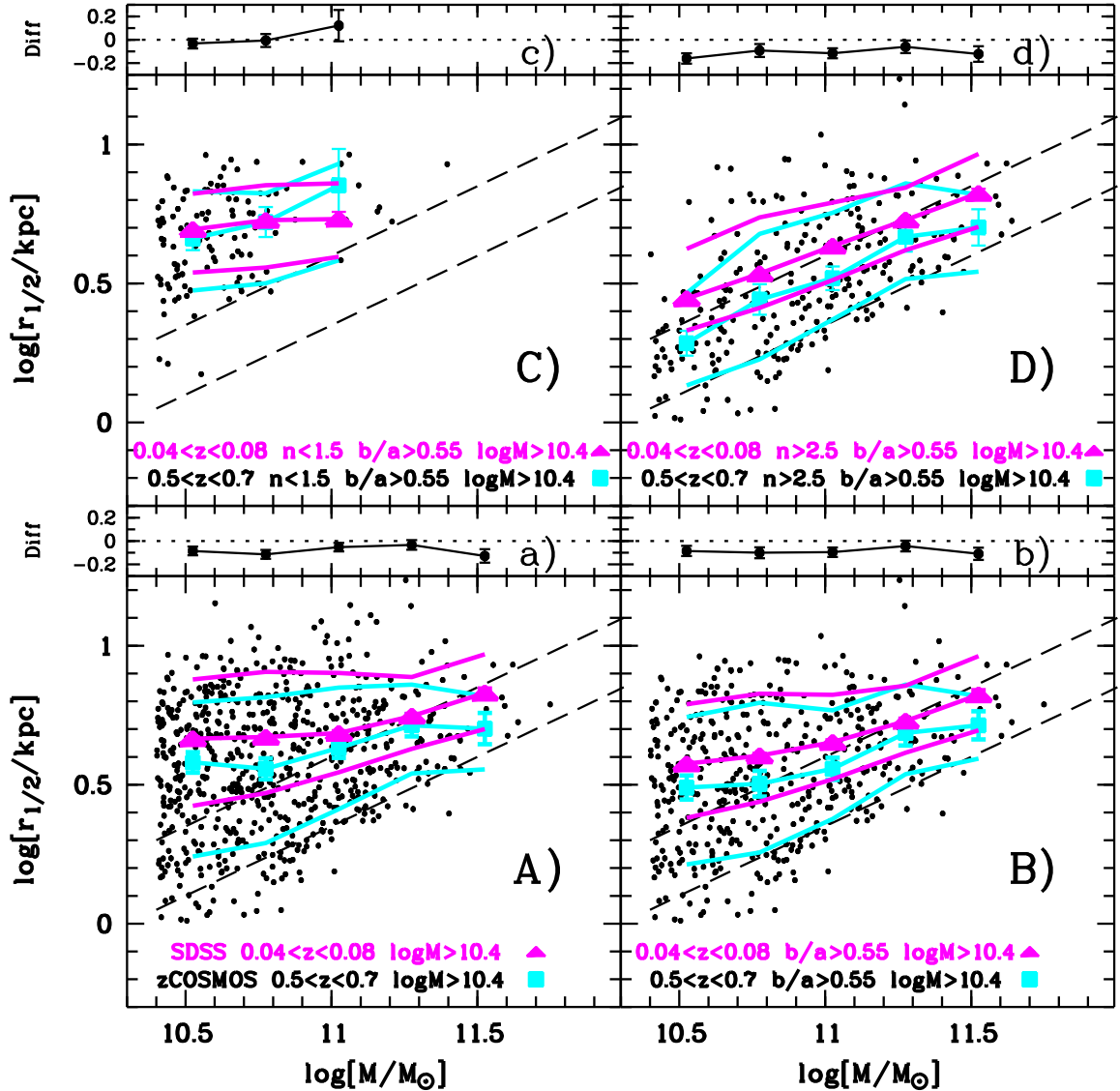}
\caption
{\label{r12Mass} \footnotesize 
The semi-major half-light radius $r_{1/2}$ in the observed ACS I-band (corresponding
roughly to rest-frame B-band for
the redshift range $0.5<z<0.9$) versus stellar mass for the zCOSMOS mass-complete sample at
$0.5<z<0.7$. Individual zCOSMOS measurements are shown as black dots, and median
values as filled cyan squares.
The median SDSS half-light radii in different mass bins derived from
g-band images are shown as filled magenta triangles, while
the 16th and 84th percentiles of
$r_{1/2}$ of the SDSS and zCOSMOS galaxies in each mass bin are
shown as solid magenta and cyan lines respectively.
The two dashed diagonal black lines show mass surface densities of $\rm{log}\Sigma_{M}=9$
(upper line), and $\rm{log}\Sigma_{M}=9.5$, respectively.
Panel a)  shows all galaxies in the mass-complete sample, panel b) 
galaxies with an axis ratio $b/a>0.55$, 
panel c)  galaxies with   $b/a>0.55$ and $n<1.5$, and panel d) galaxies
with  $b/a>0.55$ and $n>2.5$.
The respective small panels above the main ones show as filled squares the difference between the median
zCOSMOS and SDSS $r_{1/2}$  values in a given
mass bin.
For $n<1.5$ galaxies there is almost no evolution in the stellar mass - size relation 
between $z \sim 0.7$ to $z \sim 0$, while for  $n>2.5$ objects 
the average half-light radius of galaxies at a given mass is smaller by $\sim 25$\% at
$z \sim 0.7$. 
}
\end{figure}

\begin{figure}[h]
\includegraphics[width=16cm,angle=270,clip=true]{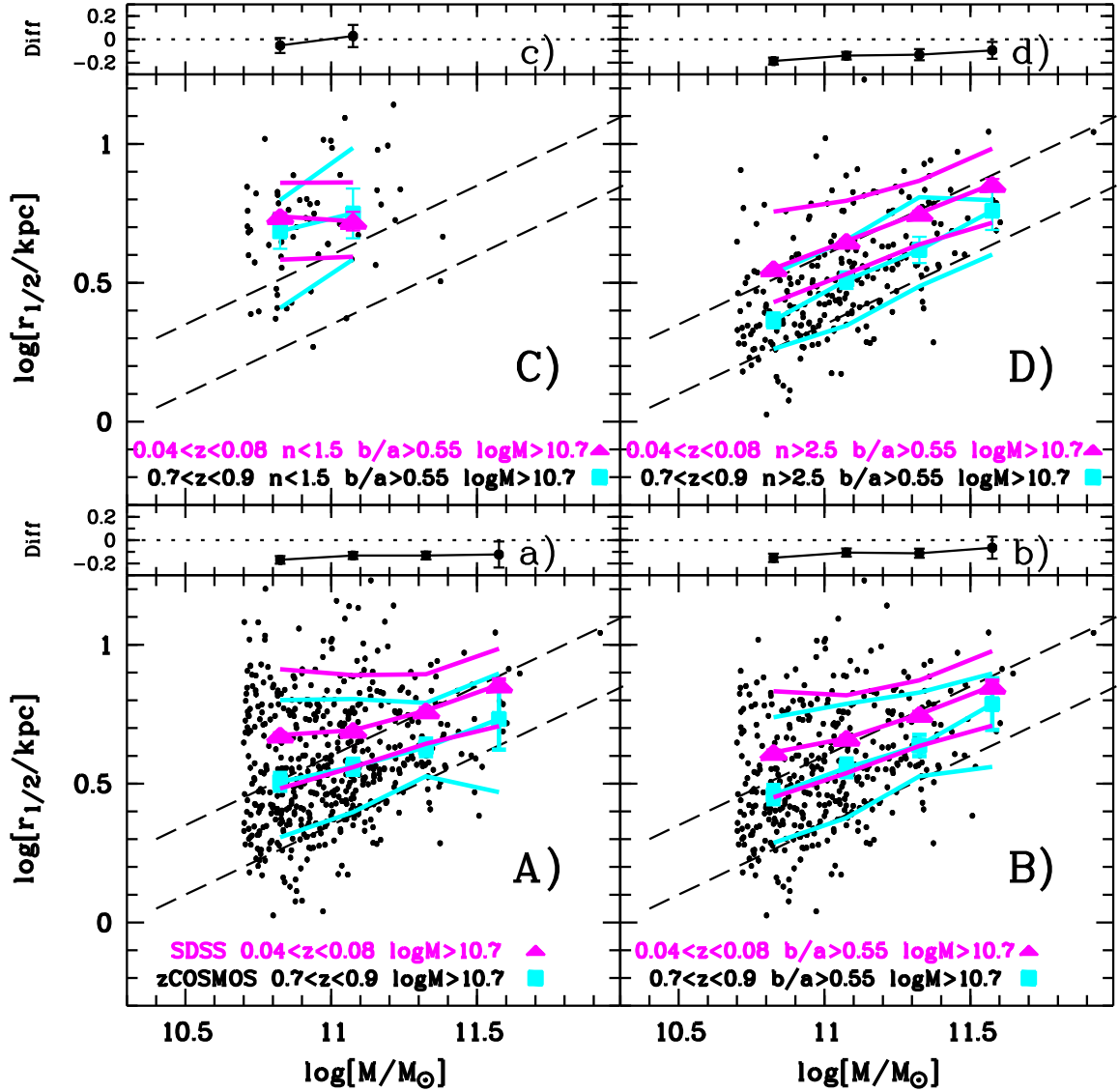}
\caption
{\label{r12Masszgt07} \footnotesize 
The semi-major half-light radius $r_{1/2}$ 
versus stellar mass for the zCOSMOS mass-complete sample at
$0.7<z<0.9$ (black dots and cyan symbols and lines) compared to SDSS
(magenta symbols and lines).
Symbols are as in Fig.\,\ref{r12Mass}, and similar trends of the stellar mass - size relation are seen
here for $0.7<z<0.9$ zCOSMOS galaxies as seen in Fig.\,\ref{r12Mass} for the slightly lower redshift
$0.5<z<0.7$ objects.
However, the
change in sizes of the early-type
galaxies (panel d) becomes even more pronounced  and this leads to a further increase in
$\Sigma_{Mtrans}$ and $\Sigma_{Mchar}$.
}
\end{figure}

In order to better understand the small shift in $\Sigma_{Mchar}$ with redshift, we have 
looked at the relationship between mass and half-light radius in both Sersic subsamples
at both SDSS and zCOSMOS redshifts.
Figure \ref{r12Mass} shows the semi-major half-light radius in the observed ACS I-band (corresponding
roughly to rest-frame B-band for
this redshift range) versus stellar mass for the zCOSMOS mass-complete sample at
$0.5<z<0.7$. Black dots are individual zCOSMOS measurements, while cyan
squares denote the median $r_{1/2}$ values in the respective
mass bin. 
The SDSS median half-light radii in different mass bins derived from
g-band images (as described in Sect. \ref{SDSSsample}) are shown as filled, magenta triangles.
The 16th and 84th percentiles of
$r_{1/2}$ in the SDSS and zCOSMOS galaxies in each mass bin are
shown as solid magenta and cyan lines respectively.
The two dashed black lines show constant mass surface densities of $log\Sigma_{M}=9$
(upper line), and $log\Sigma_{M}=9.5$, respectively.
As before, panel a) of Fig.\,\ref{r12Mass} shows all galaxies in the zCOSMOS mass sample, panel b)
galaxies with $b/a>0.55$, panel c)  galaxies with $b/a>0.55$ and $n<1.5$, and panel d) galaxies
with  $b/a>0.55$ and $n>2.5$.
The distribution of sizes of SDSS galaxies
is displaced to larger sizes at a given mass compared with zCOSMOS when galaxies of all
Sersic indices are considered (panels a and b). However, inspection of 
panels c and d of Fig.\,\ref{r12Mass} shows that this effect is almost entirely due
to galaxies with $n>2.5$. The respective stellar mass-size relation for the slightly higher redshift $0.7<z<0.9$ galaxies is
shown in Fig.\,\ref{r12Masszgt07}, and is discussed at the end of this section.

Panel c of Fig.\,\ref{r12Mass} shows that the stellar mass-size
relation of $n<1.5$ zCOSMOS galaxies at $0.5<z<0.7$ does not differ
significantly from the similar relation for
SDSS $n<1.5$ galaxies. This is
consistent with the GEMS result of \citet{barden05}. 
These authors interpreted the almost negligible evolution with time in the stellar mass - size
relation as support for an ``inside-out'' scenario.  We would interpret an ``inside-out'' scenario 
to involve a broadly constant surface mass density (whether measured at the center or at a
characteristic half mass radius) coupled with an increasing half-mass radius, i.e. an 
evolution parallel to the dashed lines in Fig.\,\ref{r12Mass}.

However, as remarked above, the SFR in these galaxies 
declines rapidly (by a factor of about six  as 
shown in panel c of
Fig.\,\ref{SSFR_Mareazlt07}) since $z \sim 0.7$. The subsequent
evolution of these $n<1.5$ zCOSMOS galaxies is described  presumably by a
declining SFR, as suggested by
the  paths of model galaxies  shown in Fig.\,\ref{SSFR_MassBC03} (solid green tracks). 
We show in panel c of Fig.\,\ref{SSFR_Mareazlt07} the possible tracks of
model galaxies  (red solid arrows) with declining SFRs and with sizes assumed to be constant
over this time period.  The mass, and thus stellar mass density 
increases by $\Delta M_{*} \leq
0.3 M_{*}$. Given that the stellar mass-size relation seen in panel c of
Fig.\,\ref{r12Mass} is almost flat, it is difficult to see much 
evidence for the required diagonal evolution in the size-mass plane.
We therefore do not believe, over this redshift interval at least, that
there is much evidence for inside out growth of disks.  This is also 
consistent with the lack of evolution in the size function of disks
\citep{lilly98,sargent07}.

In contrast, panel d in
Fig.\,\ref{r12Mass} illustrates the
difference in the stellar mass-size relation of high
Sersic index ($n>2.5$) zCOSMOS galaxies
compared with the corresponding SDSS sample. Since the masses of early-type
$n>2.5$ galaxies are not expected to evolve significantly between  $z \sim 0.7$ and $z \sim 0$ because of the low
and rapidly declining SSFRs (the two models on
the right in the left panel of Fig.\ref{SSFR_MassBC03}), we conclude that
zCOSMOS objects are on average 25\% smaller at a given mass (panel d and D in Fig.\,\ref{r12Mass}).
This change in size is similar to
but smaller than  found by \citet{trujillo07}, who  claimed $\sim 60$\% smaller $r_{1/2}$ on average compared to SDSS for objects at similar redshift with somewhat higher stellar
masses ($\rm{logM}_{*}>11$).

\begin{figure}[h]
\includegraphics[width=12cm,angle=270,clip=true]{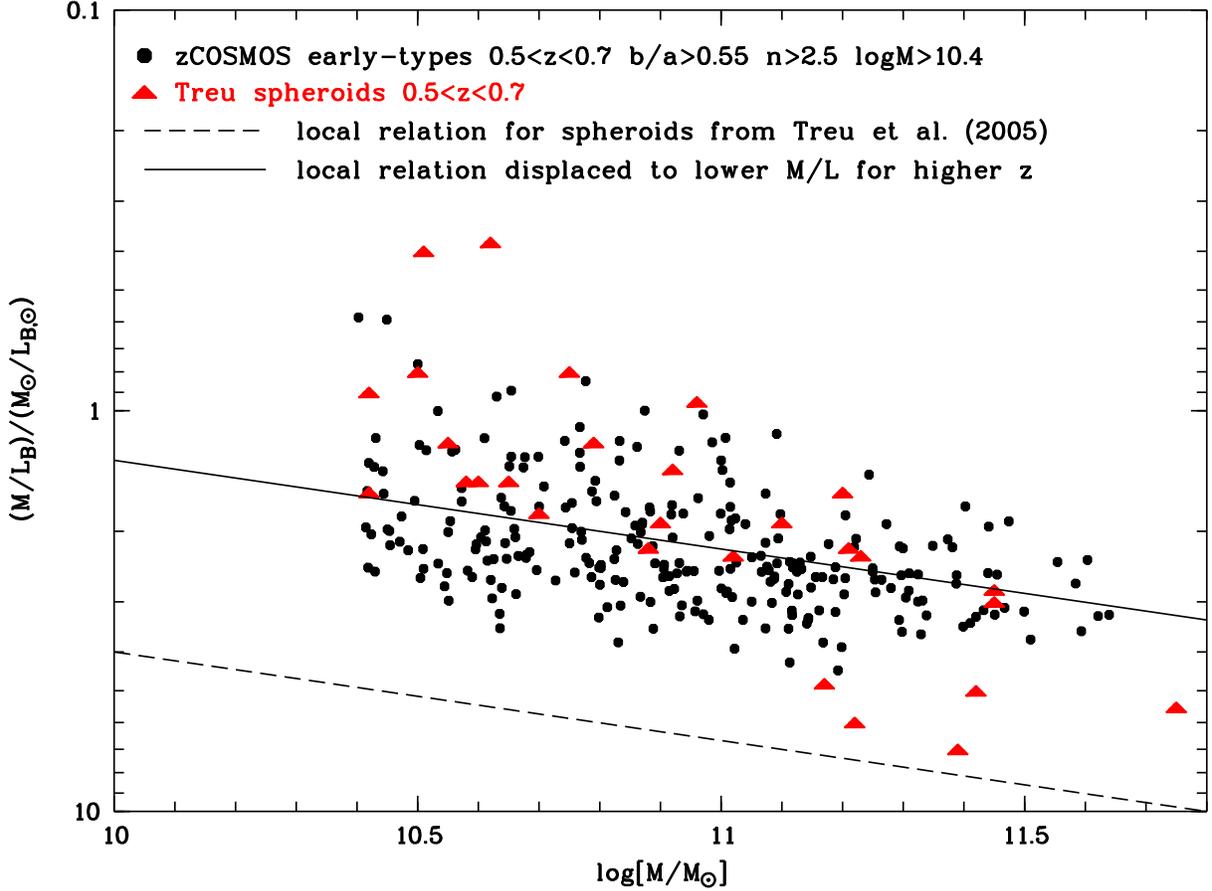}
\caption
{\label{M_MLrel} \footnotesize 
Comparison of the B-band M/L ratios for zCOSMOS galaxies with $b/a>0.55$
and $n>2.5$ (black filled circles) with the
M/L ratios obtained studying the fundamental plane for early-type (E+S0) galaxies from
Fig.\,15 of \citet{treu05}, and the local relation for early-type (spheroid)
galaxies taken from the same figure of \citet[][dashed line]{treu05}. 
 The slope of the observed mean zCOSMOS M/L-mass relation 
agrees with the slope of the local relation, as indicated by the solid diagonal line, offset 
to lower M/L, as expected for a passively evolving
population. 
Moreover, the spheroids from \citet{treu05} occupy a similar region of
the M/L-mass diagram as the zCOSMOS galaxies at $0.5<z<0.7$.
This reassures that the derived stellar
masses for $n>2.5$ zCOSMOS galaxies are reasonable, and further suggests that the changes in the stellar mass-size relation 
at $0.5<z<0.7$
in  Fig.\,\ref{r12Mass} are
due mainly to  smaller sizes (or at least smaller measured $r_{1/2}$), at a given mass, in
zCOSMOS compared with SDSS.
}
\end{figure}

%
One might worry that  stellar masses for 
$n>2.5$ zCOSMOS galaxies using Eq.\,\ref{logM} are systematically wrong.
To check the reliability of the calculated zCOSMOS masses,
we compare the B-band mass-to-light ratios of the zCOSMOS $n>2.5$ galaxies with the
M/L ratios obtained by studying the fundamental plane for early-type (E+S0) galaxies by
\citet{treu05}.
Figure \ref{M_MLrel} shows the B-band M/L ratios as a function of mass
for our zCOSMOS sample (black filled
circles), the M/L ratios of spheroids  (E+S0) galaxies at $0.5<z<0.7$ from
Fig.\,15 of \citet{treu05}, and the local relation for early-type (spheroid)
galaxies (dashed line) taken from their same figure.
The slope of the mean observed zCOSMOS  M/L-mass relation 
agrees with the slope of the local relation, as indicated by the solid diagonal line in Fig.\,\ref{M_MLrel}, which is obtained by shifting the local relation to the lower M/L expected for a passively evolving
population. 
Moreover, the $0.5<z<0.7$ spheroids from \citet{treu05} occupy a similar region of
the M/L - mass diagram as the zCOSMOS galaxies at the same redshifts, suggesting 
that our stellar
masses derived from optical colors for $n>2.5$ (early-type) galaxies are
reasonable, and further suggesting that the changes in the stellar mass-size relation 
at $0.5<z<0.7$
in  Fig.\,\ref{r12Mass} are
due mainly to  smaller sizes (or at least smaller measured $r_{1/2}$), at a given mass, in
zCOSMOS compared with SDSS.

Similar results regarding the size evolution of early-type galaxies have been seen by \citet{trujillo07}, with an even larger change 
in size, and at higher redshifts $z>1.4$ by \citet{cimatti08}.  Both these authors
advocated dry mergers as a possible mechanism for the
growth in size.
%
%
Simulations with realistic  boundary conditions by
\citet{naab07} indicated that mergers could increase the size of galaxies
while the mass  changes only slightly. In the smooth envelope
accretion scenario of \citet{naab07}, accreted stars (mainly provided by
minor mergers) form an envelope whose size increseases smoothly with time.
\citet{khochsilk06}  used a semi-analytical model of galaxy
formation to predict the redshift-size evolution of elliptical
galaxies, and their Fig.\,4 shows that  elliptical
galaxies with masses above $10^{10}M_{\odot}$ could increase their
sizes by a factor of about 1.25 from $z \sim 0.7$ to $z \sim 0$, 
consistent with our findings in panel D) and d) of Fig.\,\ref{r12Mass} for $n>2.5$ galaxies.

%

\begin{figure}[h]
\includegraphics[width=16cm,angle=270,clip=true]{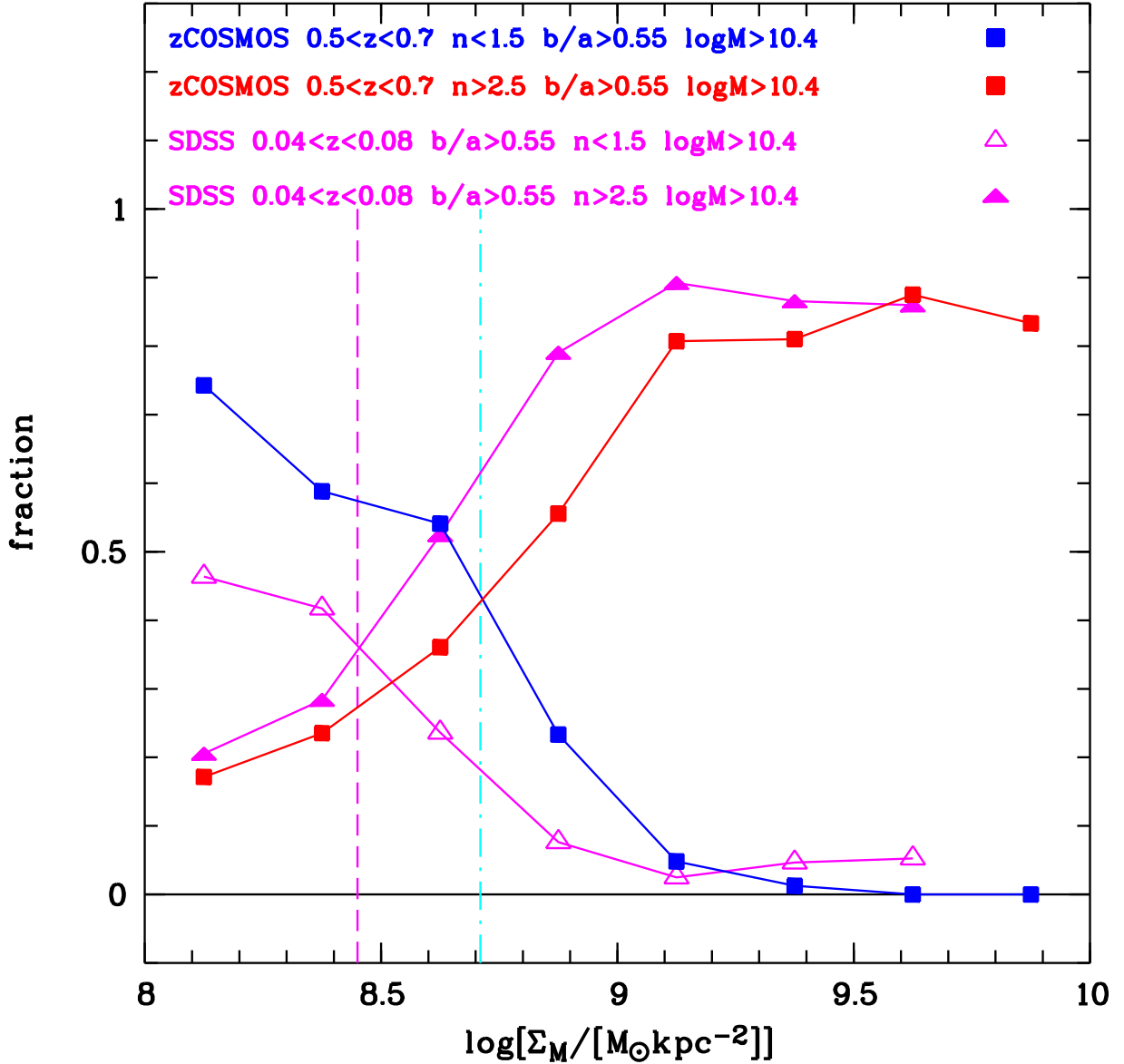}
\caption
{\label{fracngt25zlt07} \footnotesize 
The fraction of $b/a>0.55$ objects with $n<1.5$ (blue filled squares) and
$n>2.5$ (red filled squares) for zCOSMOS
galaxies at $0.5<z<0.7$ from the mass-complete sample with $\rm{logM}_{*}>10.4$.
The  respective fractions for SDSS galaxies are shown as filled  and
open  magenta triangles, for
$n>2.5$ and $n<1.5$ galaxies, respectively.
The transition surface mass density $\Sigma_{Mtrans}$ (dot-dashed cyan
vertical line) at which these curves cross 
is larger for zCOSMOS than for SDSS galaxies (dashed magenta vertical line), and gives an
explanation for the shift seen in the location of the step $\Sigma_{Mchar}$ in the median SSFR versus $\Sigma_{M}$ relation 
for zCOSMOS galaxies compared to that of the SDSS
(Figs.\,\ref{SSFR_Mareazlt07} and \ref{SSFR_Mareazgt07}).
}
\end{figure}

With these results in mind, we now return to understand 
the changing location of $\Sigma_{Mchar}$,
the break in the SSFR$-\Sigma_{M}$ relation.  Motivated by our SDSS analysis above,
we show in Fig.\,\ref{fracngt25zlt07}
the fraction of $n<1.5$ and $n>2.5$
Sersic index galaxies compared with the total number of galaxies (always with
$b/a>0.55$) in
different $\Sigma_{M}$ bins for zCOSMOS and SDSS.  The cross-over point $\Sigma_{Mtrans}$
shifts to higher surface 
mass densities, 
presumably because, while the average $\Sigma_{M}$ of low Sersic index galaxies ($n < 1.5$) 
remains 
unaltered, that of the higher Sersic index galaxies ($n > 2.5$) shifts at higher
redshifts to higher values, because of the apparent size evolution noted above.
This shift in $\Sigma_{Mtrans}$ then produces the shift in $\Sigma_{Mchar}$ 
in the SSFR$-\Sigma_{M}$ relation.

%
The behaviour of the SSFR - $\Sigma_{M}$ and stellar mass - size relations for
$0.7<z<0.9$ zCOSMOS galaxies (Figs.\,\ref{SSFR_Mareazgt07} and \ref{r12Masszgt07})
in comparison with the
equivalent SDSS samples (selected at higher masses because of the higher mass
completeness limit in zCOSMOS, see Fig.\,\ref{SSFR_MassBC03})
is similar to that of galaxies at the slightly
lower redshift of $0.5<z<0.7$, on which we have primarily focussed in the foregoing.  
However, the
change in sizes of the early-type
galaxies becomes even more pronounced at $0.7<z<0.9$ and this leads to a further increase in
$\Sigma_{Mtrans}$ and $\Sigma_{Mchar}$.


%
\section{Summary and conclusions}
%

We have studied the relationships between SSFR and stellar mass density in two mass-complete samples
of galaxies ($\rm{log}M_{*}>10.4$ for
$0.5 < z < 0.7$ and $\rm{log}M_{*}>10.7$ for $0.7 < z < 0.9$) drawn from the zCOSMOS survey, and the
equivalent mass-complete samples of SDSS objects to draw the following conclusions:

\begin{itemize}
%
%
\item
The median SSFR of SDSS
galaxies is almost independent
of $\Sigma_{M}$ for low values of $\Sigma_{M}$, but then abruptly changes at a surface mass
density $\Sigma_{Mchar} \sim 8.5$, in agreement with previous studies. 
This step-function is clearly due to the change-over of different structural types
from disk-dominated low Sersic galaxies ($n < 1.5$) to bulge-dominated high Sersic galaxies ($n > 2.5$)
as the $\Sigma_{M}$ increases.  The population mix changes over at a transition surface mass density
$log \Sigma_{Mtrans} \sim 8.45$,
which is almost identical to the characteristic surface mass density at which the SSFR 
changes, $\Sigma_{Mchar}$,
identified by \citet{kaufm06}.
\item
The shape of the SSFR$-\Sigma_{M}$ relation in zCOSMOS at $0.5 < z < 0.9$ is
very similar to that of the SDSS, with a roughly uniform increase in
the average SSFR
by a factor of 5-6 that is broadly independent of $\Sigma_{M}$ and which occurs in both
early and late type galaxies with $n > 2.5$ and $n < 1.5$ respectively. There is also
a small increase of 0.1-0.2 dex in $\Sigma_{Mchar}$ at the higher redshifts.

\item
The rise in SSFR is almost exactly the same as that seen in the overall
star-formation rate density (SFRD) of the Universe to this redshift, implying that
galaxies across the full range of $\Sigma_{M}$, and with a wide range of 
Sersic index, are all contributing, proportionally, to the increase in 
the SFRD.

The modest increase in $\Sigma_{Mchar}$ by $\sim 0.1-0.2$ dex in
zCOSMOS relative to the SDSS is naturally explained 
by differences in the size-mass relations for  disk-dominated ($n<1.5$) and bulge-dominated
($n>2.5$) galaxies. Whereas the former have a size-mass relation that does not change with redshift,
as also found by \citet{barden05}, the latter are smaller at higher
redshifts, as also seen in the
samples of \citet{trujillo07}, pushing individual galaxies to high $\Sigma_{M}$ values.  This
increases $\Sigma_{Mtrans}$, and thus $\Sigma_{Mchar}$, at the higher redshifts.

\item
The median SSFR of disk-dominated galaxies ($n < 1.5$) is almost
independent of surface mass density at both redshifts, but is about six times
higher in zCOSMOS galaxies at $z \sim 0.7$ than in SDSS.
With this strong decline in the average SSFR, the 
masses of the disk-dominated galaxies are unlikely to grow by more than $\sim 30$\% from $z \sim 0.7$ to $z \sim 0$.  Coupled
with the observed flat relation between size and stellar mass, this makes the evidence for 
``inside-out'' growth of disks weak, at least 
over this redshift range \citep[c.f.][]{barden05}.
%
%
\item
The upturn in the median SSFR for $n>2.5$
galaxies at lower $\Sigma_{M}$ is evidently due to galaxies with a significant
disk component that have Sersic index $n>2.5$ due to a large dominant bulge.
Their exclusion results in a median  SSFR that is independent of
$\Sigma_{M}$ also for the remaining $n>2.5$ zCOSMOS galaxies, but is
about a factor of five to six lower 
than for the $n<1.5$ zCOSMOS galaxies.
\end{itemize}

Putting these together we arrive at the following main observation concerning the evolution of
relatively massive galaxies since $z \sim 1$.  In parallel with the 
evidence that, at all redshifts, the mean SSFR {\it within} a given population (either disk-dominated with $n < 1.5$ 
or bulge-dominated with $n > 2.5$) is {\it independent} of $\Sigma_{M}$, and that
the observed SSFR$-\Sigma_{M}$ step-function relation is due, at all redshifts, to the changing mix of
disk-dominated and bulge-dominated galaxies as $\Sigma_{M}$ increases and the 
strong difference in  the average SSFR between disks and bulges, we find that
the {\it increase} in SSFR  with redshift is also {\it independent} of $\Sigma_{M}$ 
and also 
of Sersic index $n$.  The increase matches that of the global star-formation rate density of the Universe as a whole,
suggesting that all types of galaxies are participating in the increase in SFR.

The conclusion is that the internal build-up of stellar mass in a galaxy is not strongly affected 
by $\Sigma_{M}$, beyond the overall structural role of $n$ (which clearly has a very large effect 
on the average SSFR, no doubt through the different structural components of disk and spheroid).
The (relative) mean increase in SSFR back to $z \sim 1$ not only does not depend on $\Sigma_{M}$,
it does not even seem to depend on $n$.

\acknowledgments
We would like to thank the anonymous referee for his or her
suggestions and comments.
We also want to thank Mariano Ciccolini and  Monique Aller for their
help for testing the reliability of COSMOS size measurements.
C.M. acknowledges support from the Swiss National Science Foundation.

%
%

%
%


\clearpage


\clearpage


\clearpage


\clearpage

\begin{appendix}
\section{zCOSMOS vs. SDSS size measurements}
\label{sizetest}

\begin{figure}[h]
\includegraphics[width=8cm,angle=270,clip=true]{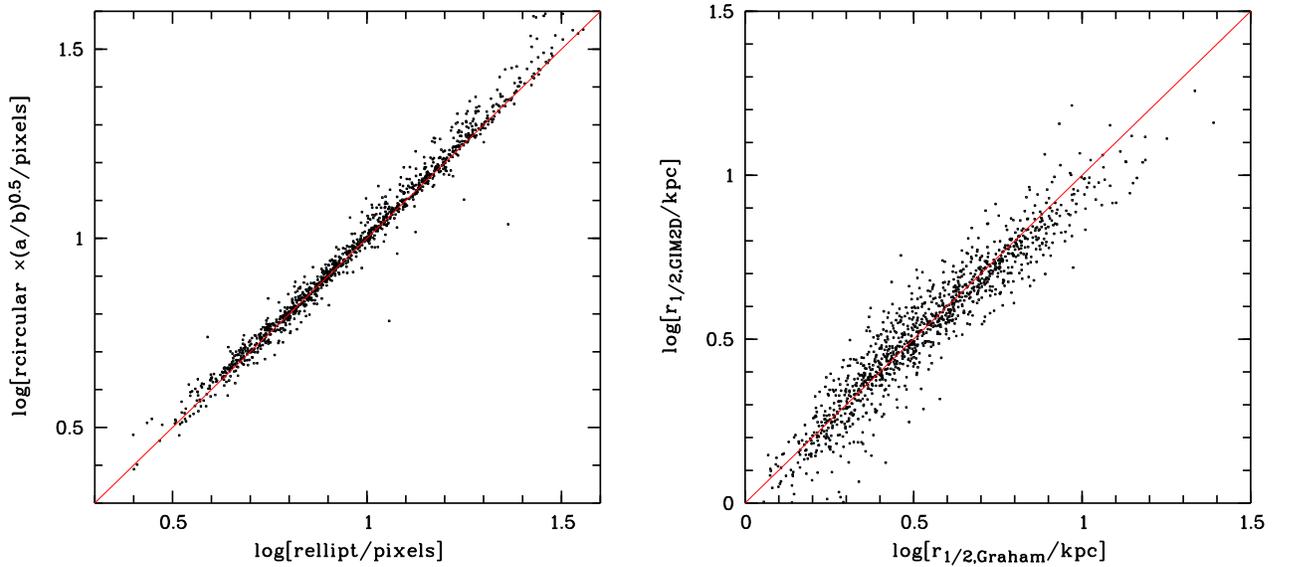}
\caption
{
\label{SizesVergl}
The left panel shows a good agreement (offset 0.009 dex, RMS 0.03 dex)
between the COSMOS sizes (semi-major axis) computed directly from the ACS images using an elliptical aperture, and the sizes
computed using  circular apertures and multiplying them by $\sqrt{a/b}$.
The right panel shows also a quite good agreement (offset 0.008 dex,
RMS 0.07 dex) between the sizes (semi-major axis) computed with GIM2D,
and the sizes computed starting from the circular Petrosian radius containing
50\% of the Petrosian flux and applying
equation 6 of \citet{graham05} and the $\sqrt{a/b}$ correction to
compute the semi-major half-light radii (the same method as applied to
SDSS and described in Sect.\,\ref{SDSSsample} a).
}
\end{figure}

It might be a concern that there are significant systematic differences
between the size measurement methods for zCOSMOS (as described in
Sect.\,\ref{morphpar}) and SDSS (as described in Sect.\,\ref{SDSSsample}a).
To check  the consistency of our size measurements for the two samples 
we used the ACS images of the mass-complete sample of $\sim 1200$
zCOSMOS galaxies at $0.5<z<0.9$ to test two issues: i) are the sizes
determined using a circular aperture (as in the case of SDSS) with a
correction of 
$\sqrt{a/b}$ consistent with the sizes determined using elliptical
apertures (as in GIM2D applied to zCOSMOS);
and ii) are the GIM2D  half-light radii consistent with a derivation
of half-light radii starting from the Petrosian radius containing 50\%
of the Petrosian flux and applying the corrections done for the SDSS
sample described in
Sect.\ref{SDSSsample}a ? 
The results are shown in the two panels of 
Fig.\,\ref{SizesVergl}, and show that there are no significant
systematic differences between the two size measurement
methods in zCOSMOS and SDSS.

\end{appendix}


\end{document}